%% file: all7.tex
\documentclass[12pt, epsf]{article}
\input eps.tex

\usepackage{latexsym}
\usepackage{amsmath}
\usepackage{amssymb}
\usepackage{amsfonts}
\usepackage{times}
\usepackage{graphicx}
\usepackage{epsfig}
\usepackage{array}
\usepackage{flafter}

\newcommand{\Ex}[1]{Example~\ref{ex#1}}
\newcommand{\ex}[1]{example~\ref{ex#1}}

\newtheorem{ExampleDef}{Example}[section]

\newcommand{\Example}[3]{
  \begin{list}{}{
      \setlength{\leftmargin}{1em}} 
    \item                           
    \small                          
    \begin{ExampleDef} \rm          
      {\bf \hspace{-1ex}: #1}       
      #2                                
      \hfill {\large \boldmath $\Box$}  
      \label{ex#3}                      
    \end{ExampleDef}
  \end{list}}

\begin{document}

\begin{flushright}
{\small NSL-980401. October, 2009.}
\end{flushright}

\begin{center}
{\Large \bf Shaping state and time-dependent convergence rates in
  non-linear control and observer design \par}
\vspace{1.5em}
{\large Winfried Lohmiller and Jean-Jacques E. Slotine \par}
{Non-Linear Systems Laboratory\\
Massachusetts Institute of Technology \\
Cambridge, Massachusetts, 02139, USA\\
{\sl wslohmil@mit.edu, jjs@mit.edu} \par}
\vspace{3em}
\end{center}

\begin{abstract}

This paper derives for non-linear, time-varying and feedback
linearizable systems simple controller designs to achieve specified
{\it state-and time-dependent} complex convergence rates. This
approach can be regarded as a general gain-scheduling technique with
global exponential stability guarantee. Typical applications include
the transonic control of an aircraft with strongly Mach or
time-dependent eigenvalues or the state-dependent complex eigenvalue
placement of the inverted pendulum.

As a generalization of the LTI Luenberger observer a dual observer
design technique is derived for a broad set of non-linear and
time-varying systems, where so far straightforward observer techniques
were not known. The resulting observer design is illustrated for
non-linear chemical plants, the Van-der-Pol oscillator, the discrete
logarithmic map series prediction and the lighthouse navigation
problem.

These results \cite{Lohm6} allow one to shape globally the state-
and time-dependent convergence behaviour ideally suited to the
non-linear or time-varying system. The technique can also be used to
provide analytic robustness guarantees against modelling uncertainties.

The derivations are based on non-linear contraction theory
\cite{Lohm1}, a comparatively recent dynamic system analysis tool
whose results will be reviewed and extended.
\end{abstract}

\section{Introduction}

Non-Linear contraction theory \cite{Chung, Lohm0, Lohm1, Lohm2, Lohm3,
Lohm4, Lohm5, Lohm6, Lohm7, jjs01, wei} is a comparatively recent
dynamic analysis and design tool based on an exact differential
analysis of convergence.  Similarly to chaos theory, contraction
theory converts a non-linear stability problem into a LTV (linear
time-varying) first-order stability problem by considering the
convergence behaviour of neighbouring trajectories. Global convergence
can be concluded since a chain of neighbouring converging trajectories
also implies convergence over a finite distance. A brief summary of
contraction theory is given in section \ref{contraction}.

Whereas chaos \cite{Chaos} and LTV theory \cite{Kailath} in section
\ref{chaosLTV} compute {\it numerically} the transition matrix and
hence the {\it time-averaged} convergence rates in form of the
Lyapunov exponents, contraction theory provides explicit {\it
analytical} bounds on the {\it instantaneous} convergence or
contraction rate. Note that the incremental stability approach in
\cite{Angeli} extends these instantaneous analytical contraction rates
to the {\it integrated} convergence of neighbouring trajectories.

Since contraction theory assesses the convergence of {\it
all neighbouring} trajectories to each other, it is a stricter stability
condition than Lyapunov convergence, which only considers 
convergence to an {\it equilibrium point}. It is this difference
which enables observer or tracking controller designs, which do not
converge to an equilibrium point. Also, contraction convergence results
are typically exponential, and thus stronger than those based on 
most Lyapunov-like methods.

So far contraction analysis relied in section \ref{contraction} on
finding a suitable {\it metric} to bound the contraction rate of a
system. Depending on the application, the metric may be trivial
(identity or rescaling of states), or obtained from physics (say,
based on the inertia matrix in a mechanical system), combination of
simpler contracting subsystems \cite{Lohm1}, semi-definite programming
\cite{Lohm4}, sums-of-squares programming \cite{Parrilo}, or recently
contraction analysis of Hamiltoninan systems \cite{Lohm7}.

This paper \cite{Lohm6} shows that the computation of the metric may
be largely simplified or indeed avoided altogether by extending the
{\it first-order} exact differential analysis to the placement of
state-or time-dependent contraction rats of {\it
$n$-th-order} ($n \ge 1$) continuous systems in controllability form
$$ 
{\bf x}^{(n)} = {\bf f}({\bf x}, ..., {\bf x}^{(n-1)}, t) + {\bf
G}({\bf x}, ..., {\bf x}^{(n-1)}, t) {\bf u}
$$
with $N$-dimensional position ${\bf x}$, $M$-dimensional control input
${\bf u}$ and time $t$. In addition a dual observer design is derived
for smooth $n$-th order dynamic systems in observability form
$$
{\bf x}^{(n)} = {\bf a}_1^{(n-1)}({\bf x}, t) + {\bf
a}_2^{(n-2)}({\bf x}, t) + ...  + {\bf a}_n({\bf x}, t)
$$
with $M$-dimensional measurement ${\bf y}({\bf x}, t)$, $N$-dimensional
position ${\bf x}$, $N$-dimensional non-linear plant
dynamics ${\bf a}_j({\bf x}, t)$ and time $t$.

A similar method to place state- and time-dependent contraction rates
will also be derived for the corresponding discrete controllability form 
$$ 
{\bf x}^{i+n} = {\bf f}({\bf x}^i, ..., {\bf x}^{i+n-1}, i) + {\bf
G}({\bf x}^i, ..., {\bf x}^{i+n-1}, i) {\bf u}^i
$$
with $N$-dimensional position ${\bf x}^i$, $M$-dimensional control
input ${\bf u}^i$ and time index $i$. In addition a dual observer
design is derived for smooth $n$-th order dynamic systems in
observability form
$$
{\bf x}^{i+n} = {\bf a}_1^{(+n-1)}({\bf x}^i, i) + {\bf
a}_2^{(+n-2)}({\bf x}^i, i) + ... + {\bf a}_n({\bf x}^i, i) 
$$
with $M$-dimensional measurement ${\bf y}({\bf x}^i, i)$,
$N$-dimensional position ${\bf x}^i$, $N$-dimensional
non-linear plant dynamics ${\bf a}_j({\bf x}^i, i)$ and time index
$i$. The superscript $(+j)$ implies now and in the following that the
function is mapped $j$ times in the future.

The following example illustrates the relation of this paper to the
design of standard LTI controllers and shows that for non-linear,
time-varying systems, stable convergence is not quantified by the linearized
eigenvalues, but by the contraction rates as defined in this paper.

\Example{}{Consider the simplified A/C angle-of-attack dynamics
$$
\ddot{\alpha} + D(q_c, Ma) \dot{\alpha} + K(q_c, Ma) \alpha = G(q_c,
Ma) u
$$
with angle-of-attack $\alpha$, dynamic pressure $q_c(t)$, Mach number
$Ma(t)$ and control input $u$.  In a generalization of feedback
linearization let us now schedule the complex eigenvalues
$\lambda_1(Ma, q_c)$ and $\lambda_2(Ma, q_c)$ with $Ma(t)$ and
$q_c(t)$ to reflect this strong non-linear plant dependence in the A/C
controller. This yields the hierarchical or cascaded system with $z_1
= \alpha$
\begin{eqnarray}
\dot{z}_1 &=& \lambda_1 z_1 + z_2 \nonumber \\
\dot{z}_2 &=& \lambda_2 z_2 + u_d(t) \nonumber 
\end{eqnarray}
which implies the control input $u$
\begin{eqnarray}
\ddot{\alpha} &=& - D(q_c, Ma) \dot{\alpha} - K(q_c, Ma) \alpha +
G(q_c, Ma) u \nonumber \\
&=& u_d(t) + (\lambda_1 + \lambda_2) \dot{\alpha} - ( \lambda_1
\lambda_2 - \dot{\lambda}_1) \alpha \nonumber
\end{eqnarray}
where $\lambda_1$ and $\lambda_2$ have to be chosen such that $u$
stays real. The key difference to standard gain-scheduling techniques
(see e.g. \cite{Lawrence}) is the term $\dot{\lambda}_1(Ma,
q_c)$. Only with this term exponential convergence with $\lambda_1$
and $\lambda_2$ to the desired trajectory, defined by $u_d(t)$, is
guaranteed.}{accontroller}
A major point of this paper will be the extension of
\ex{accontroller} to the observer and controller design of complex
state- and time-dependent contraction rates, considering the
time-derivatives of the contraction rates to make the analysis
correct.

In section \ref{contcontroller} state- and time-dependent contraction
rates are ``placed'', as a generalization of standard feedback
linearization methods (see e.g. \cite{Isidori},\cite{flatness} or
\cite{Reboulet}). The generalization is that we can choose state- or
time-dependent contraction rates $\lambda_j({\bf z}_j, t)$ to simplify
$u$, to handle only piece-wise controllable systems (under-actuacted or
intermittently controlled systems), as e.g. in the inverted pendulum
or in legged locomotion, or simply to achieve state- or time-dependent
system performance.  In contrast to standard gain-scheduling
techniques (see e.g. \cite{Lawrence}, \cite{Mracek}, \cite{Reboulet})
global exponential stability guarantees of the state- and
time-dependent contraction rates are still given.

Section \ref{contobserver} derives a corresponding non-linear observer
design. It extends the LTV Luenberger observer of constant eigenvalues
in \cite{Zeitz} to higher-order non-linear systems with designed state-
and time-dependent desired contraction rates.

The corresponding stability analysis of a given higher-order system is
presented in section \ref{continuoushigher}. This technique also
allows to bound analytically the robustness of a given controller and
observer design with respect to modelling uncertainties.

Section \ref{discontroller}, \ref{disobserver} and
\ref{discretehigher} extend the controller and observer design
technique to the discrete case. We e.g. assess the stability of a
non-linear price/demand dynamics, design an observer for the logistic
map problem or derive a simple non-linear global observer for the
standard bearings-only or lighthouse problem, of navigating a vehicle
using only angular measurements with respect to a fixed point in space
\cite{Bekris}. The algorithm is non-linear but very simple. It is new
to our knowledge, and provides explicit global convergence
guarantees. The algorithm's stochastic version should serve as a
simpler and ``exact'' alternative to approaches based on linearization
and the extended Kalman filter, both in the pure bearings-only problem
and as part of more complex questions such as simultaneous
localization and mapping (SLAM).

Concluding remarks are offered in section \ref{Conclusion}.

\section{Relation to chaos theory} \label{chaosLTV}

Contraction theory and chaos make extensive use of {\it virtual
displacements}, which are differential displacements at fixed time
borrowed from mathematical physics and optimization theory. Formally,
if we view the $N$-dimensional position ${\bf x}$ of the system at
time $t$ as a smooth function of the initial condition ${\bf x}^o$ and
of time, $\ {\bf x} = {\bf x}({\bf x}^o ,t)\ $ we get $\ \delta {\bf
x} = \Phi(0, t) \ \delta {{\bf x}^o}$ with the transition matrix
$\Phi(0, t) = \frac{\partial {\bf x}}{\partial {\bf x}^o}({\bf
x}^o(0), 0, {\bf x}(t), t)$.

Consider now an $N$-dimensional, non-linear, time-varying discrete
system 
$$ 
{\bf x}^{i+1} = {\bf f}({\bf x}^i, i)
$$ 
The convergence behaviour of neighbouring trajectories is then given by
the discrete virtual dynamics
$$
\delta {\bf x}^{i+1} = {\bf F} \delta {\bf x}^i
$$
with ${\bf F} = \frac{\partial {\bf f}}{\partial {\bf x}^i}({\bf
  x}^i, i)$. The transition of any virtual displacement from $0$ to
$i$ is then given by
$$
\delta {\bf x}^i = \Phi(0, i) \delta {\bf x}^o
$$
with the transition matrix
\begin{equation}
\Phi(0, i) = {\bf F}^{(-1)} ... {\bf F}^{(-i)}
\label{eq:disctransition}
\end{equation}
where the superscript $(+j)$ implies that the function is mapped $j$
times in the future.

Consider now an $N$-dimensional, non-linear, time-varying continuous
system 
\begin{equation} 
\dot{\bf x} = {\bf f}({\bf x}, t) \nonumber
\end{equation}
The convergence behaviour of neighbouring trajectories is then given 
by the continuous virtual dynamics
$$
\delta \dot{\bf x} = {\bf F}({\bf x}, t) \delta {\bf x}
$$
with ${\bf F} = \frac{\partial {\bf f}}{\partial {\bf x}}({\bf
  x}, t)$. The transition of any virtual displacement from $0$ to
$t$ is then given by
$$
\delta {\bf x} = \Phi(0, t) \delta {\bf x}^o
$$
with the transition matrix
\begin{equation}
\Phi(0, t) = {\bf I} + \int_{0}^t {\bf F}(\tau_1) d \tau_1 +
\int_{0}^t {\bf F}(\tau_1) \int_{0}^{\tau_1} {\bf F}(\tau_2) d
\tau_2 d \tau_1 + ... \label{eq:transition}
\end{equation}
which is equivalent to $e^{\int_o^t {\bf F} dt}$ for a diagonal
Jacobian ${\bf F}$.

The Lyapunov components (see e.g. \cite{Chaos}) simply correspond to
the N'th square root of the singular values of $\Phi(0, t)$ or
$\Phi(0, i)$. Note that the coordinate invariance of this dynamics
under smooth coordinate transformations is shown for $i, t \rightarrow
\infty$ in \cite{Chaos}. The major problem of chaos theory is that in
general the above has to be computed numerically.

What is new in contraction theory is that the transition matrices
above can be exponentially over/under-bounded in analytical form. This
will be shown in the following section in Theorem \ref{th:theoremF}
and \ref{th:theoremFdis}.

\section{First-order contraction theory} \label{contraction}

Consider now an $N$-dimensional, non-linear, time-varying, complex
continuous system 
\begin{equation} 
\dot{\bf x} = {\bf f}({\bf x}, t) \nonumber
\end{equation}
The convergence behaviour of neighbouring trajectories is then given 
by the continuous virtual dynamics
$$
\delta \dot{\bf x} = \frac{\partial {\bf f}}{\partial {\bf x}}({\bf x},
t) \delta {\bf x}
$$
Introducing a general complex $N$-dimensional virtual displacement
$\delta {\bf z} = {\bf \Theta}({\bf x}, t) \delta {\bf x}$ leads to
the general virtual dynamics
\begin{equation}
\frac{d}{dt} \delta {\bf z} \ = \ {\bf F} \delta {\bf z} \nonumber
\end{equation}
with complex ${\bf F} \ = \ \left(\dot{\bf \Theta} + {\bf \Theta}
\frac{\partial {\bf f}} {\partial {\bf \bf x}} \right){\bf
\Theta}^{-1}$. The rate of change of a differential length
$\delta s = \sqrt{ \delta {\bf z}^{\ast T} \delta {\bf z} } \ge 0$
can now be bounded by
$$
\lambda_{\min} \delta s \le \frac{d}{dt} \delta s = \frac{ \delta {\bf
z}^{\ast T} \left( {\bf F}^{\ast T} + {\bf F} \right) \delta {\bf z}
}{2 \delta s} \le \lambda_{\max} \delta s
$$
where $\lambda_{\max} (\lambda_{\min})$ is the largest (smallest)
eigenvalue of the Hermitian part of ${\bf F}$.

Recall that a complex square matrix ${\bf A}$ is said to be {\it
Hermitian} if ${\bf A}^T = {\bf A}^*$, where $^T$ denotes matrix
transposition and $^*$ complex conjugation. The {\it Hermitian part}
of any complex square matrix ${\bf A}$ is the Hermitian matrix $1/2
({\bf A} + {\bf A}^{*T})$ . All eigenvalues of a Hermitian matrix are
{\it real} numbers.  A Hermitian matrix ${\bf A}$ is said to be {\it
positive definite} if all its eigenvalues are strictly positive $ - $
this implies in turn that for any non-zero real or complex vector
${\bf x}$, one has ${\bf x}^{*T}{\bf A}{\bf x} > 0$.

Let us now define a finite distance $s = \min_s \int_{{\bf x}(s) = {\bf
x}_1}^{{\bf x}_2} \delta s \ge 0$ between two arbitrary trajectories
${\bf x}_1$ and ${\bf x}_2$ of the dynamics as the minimum
path integral over all connecting paths $s$ \cite{Lovelock}. 
The rate of change of a finite length can now be bounded by
$$
\lambda_{\min} s \le \dot{s} = \min_s \int_{{\bf x}(s)={\bf
x}_1}^{{\bf x}_2} \frac{d}{dt} \delta s \le \lambda_{\max}
s
$$
where $\lambda_{\max} (\lambda_{\min})$ is the largest (smallest)
eigenvalue of the Hermitian part of ${\bf F}$ along the path $s$.

The basic theorem of contraction analysis \cite{Lohm1, Lohm2} can
hence be stated as

\newtheorem{theorem}{Theorem}
\begin{theorem} 
Consider the deterministic system $ \ \dot{\bf x} = {\bf f}({\bf x},t)
\ $, where ${\bf f}$ is a differentiable nonlinear complex function of
${\bf x}$ within $C^N$.

\noindent Any trajectory ${\bf x}_1$ with a distance $s = \min_s
\int_{ {\bf x}(s) = {\bf x}_1 }^{\bf x_2} \sqrt{\delta {\bf x}^{\ast
T} {\bf M} \delta {\bf x}} \ge 0$ to a given other trajectory ${\bf
x}_2$ in a metric ${\bf M(x}, t) = {\bf \Theta} ({\bf x}, t)^{\ast T}
\ {\bf \Theta}({\bf x}, t) \ge 0$ exponentially converges to ${\bf
x}_2$
within the bounds
\begin{equation}
\lambda_{\max} \ge \frac{\dot{s}}{s} \ge \lambda_{\min}
\label{eq:exbound}
\end{equation}

$\lambda_{\max}$ ($\lambda_{\min}$) is defined as the largest (smallest)
eigenvalue of the Hermitian part of the generalized
Jacobian
$$
{\bf F} \ = \ \left(\dot{\bf \Theta} + {\bf \Theta} \frac{\partial
{\bf f}} {\partial {\bf \bf x}} \right){\bf \Theta}^{-1}
$$
in the ball of radius $s$ around ${\bf x}_2$.

\noindent The system is said to be contracting (diverging) for
uniformly negative $\lambda_{\max}$ (uniformly positive
$\lambda_{\min}$). The system is said to be semi-contracting
(semi-diverging) for negative $\lambda_{\max}$ (positive
$\lambda_{\min}$) and indifferent for $\lambda_{\max} = \lambda_{\min}
= 0$.

\noindent For a u.p.d. and bounded metric also the distance $\min_s
\int_{{\bf x}(s)={\bf x}_1}^{\bf x_2} \sqrt{\delta {\bf x}^{*T} \delta
{\bf x}}$ converges uniformly exponentially with the rates above,
where however initial overshoots can occure.

Note that the region of convergence of two arbitrary trajecctories
with distance dynamics $\dot{s}/s$ in (\ref{eq:exbound}) can be
extended beyond the contracting region with Lyapunovs direct method
for the specific case that explicite orthonormal Cartesian coordinates
${\bf z}({\bf x}, t)$ with dimension $\ge N$ exist as
\begin{equation}
\frac{\dot{s}}{s} = \frac{ \Re \left( \left(\dot{\bf z}_1 - \dot{\bf
z}_2 \right)^{\ast T} \left({\bf z}_1 - {\bf z}_2 \right) \right)} {
\left( {\bf z}_1 - {\bf z}_2 \right)^{\ast T} \left({\bf z}_1 - {\bf
z}_2 \right) } \label{eq:dss}
\end{equation}
\label{th:theoremF}
\end{theorem}
\begin{figure}
\begin{center}
\epsfig{figure=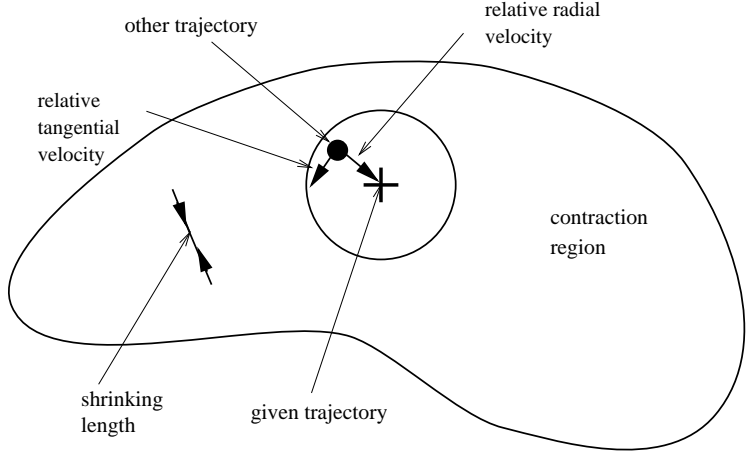,height=60mm,width=100mm}
\end{center}
\caption{Convergence of two trajectories with finite distance}
\label{fig:ball}
\end{figure}

Note that the theorem above also applies to non-differentiable ${\bf
f}$ if $\lambda_{max}$ and $\lambda_{min}$ are defined over any limit
$lim_{\Delta {\bf \bf x} \rightarrow 0}\frac{\Delta {\bf f}} {\Delta
{\bf \bf x}}$ instead of the term $\frac{\partial {\bf f}} {\partial
{\bf \bf x}}$.

Note that for a semi-contracting system (i.e. with negative semi
definite ${\bf F}$) we can conclude on asymptotic convergence if the
indefinite subspace of the symmetric part of ${\bf F}$ becomes
negative-definite in one of the higher time-derivatives of $\delta
{\bf x}^T \delta {\bf x}$ before it eventually becomes positive
definite since $\delta {\bf x}$ cannot get stuck as long as it is
unequal zero.

It can be shown conversely that the existence of a uniformly positive
definite metric with respect to which the system is contracting is
also a necessary condition for global exponential convergence of
trajectories.  In the linear time-invariant case, a system is globally
contracting if and only if it is strictly stable, with ${\bf F}$
simply being a normal Jordan form of the system and ${\bf \Theta}$ the
coordinate transformation to that form.


The following example shows how for complex systems the contraction
region of neighbouring trajectories and the region of convergence of
trajectories with a finite distance can be computed with Theorem
\ref{th:theoremF}:
\Example{}{Let us now schedule non-linear complex contraction rates
  for a second-order system by requiring the first-order complex
  dynamics
\begin{equation}
\dot{z} = - \frac{1}{2} z^2 - 2 z + u_d(t) \label{eq:zz2z}
\end{equation}
with complex contraction rate $\lambda = -z - 2$ of Theorem
\ref{th:theoremF}. In principle any differentiable complex
function can be used here to schedule the state-dependent complex
contraction rates as we want. 

The convergence rate of an arbitrary trajectory $z_1$ to another
trajectory $z_2$ is
\begin{equation}
\frac{\dot{s}}{s} = \frac{ \Re \left(\left(\dot{z}_1 - \dot{z}_2
\right)^{\ast} \left(z_1 - z_2 \right) \right)} { (z_1 - z_2)^{\ast}
(z_1 - z_2) } = \Re ( - \frac{1}{2} (z_1+z_2) - 2 )
\end{equation}
according to (\ref{eq:dss}) Theorem \ref{th:theoremF}. This region of
convergence is naturally larger then the contraction region
$\Re(\lambda) \le 0$.

The complex dynamics is illustrated in figure \ref{fig:quadratic} for
$u_d=0$. We can see that the $\Re(\lambda)$ decreases to the right.
We find exactly two equilbrium points at $z_1 = 0$ and $z_2 =-4$ with
constant distance $\frac{\dot{s}}{s} = 0$.
\begin{figure}
\begin{center}
\includegraphics[scale=0.4]{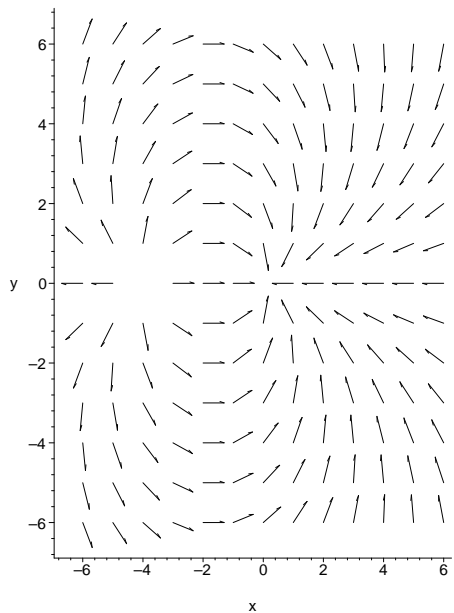}
\end{center}
\caption{Quadratic complex state space dynamics}
\label{fig:quadratic}
\end{figure}

The complex dynamics is with $x = \Re(z)$ and $y = \Im(z)$ equivalent to
\begin{eqnarray}
\dot{x} &=& -\frac{1}{2} x^2 + \frac{1}{2} y^2 - 2 x + u(t) \nonumber
\\
\dot{y} &=& -x y - 2 y \nonumber
\end{eqnarray}
with corresponding Jacobian
$$
\left( \begin{array}{cc} - x - 2 & y \\ - y & - x- 2 \end{array}
\right) 
$$
that is contracting with $\Re(\lambda)=-x-2$. 

Hence the corresponding real second-order plant dynamics of
(\ref{eq:zz2z}) is 
$$
\ddot{x} = -(6+3x) \dot{x} - x^3 - 6 x^2 - 8 x + (2 x + 4) u_d(t) +
\dot{u}_d
$$
to which the same convergence results apply.
}{complexquadratic}

For the general $N$-dimensional continuous case contraction theory
\cite{Lohm0, Lohm1} can be regarded as time-varying, complex
generalization of \cite{Hart, Kraso, Opial, Seif, Lew} with given
exponential convergence rate. In addition the introduction of the
virtual displacements in \cite{Lohm0, Lohm1} lead to a generalization
of the well-established stability and design principles of LTI systems
(see e.g. \cite{Kailath}) to the general non-linear and time-varying
case. This lead to the practical controller or observer designs in
\cite{Chung, Lohm0, Lohm1, Lohm2, Lohm3, Lohm4, Lohm6, Lohm7, wei} and
serves as a basis for this paper.

An appropriate metric to show that the system is contracting may be
obtained from physics, combination of contracting
subsystems~\cite{Lohm1}, semi-definite programming~\cite{Lohm4}, or
sums-of-squares programming~\cite{Parrilo}. The goal of this paper is
to show that the computation of the metric may be largely simplified
or avoided altogether by considering the system's {\it higher-order}
virtual dynamics.

Similarly, for a discrete system we can state 
\begin{theorem} 
Consider the deterministic system $ \ {\bf x}^{i+1} = {\bf f}({\bf
x}^i, i) \ $, where ${\bf f}$ is a smooth non-linear complex function
of ${\bf x}^i$ within $C^N$.

Any trajectory ${\bf x}_1^i$ with a distance $s^i = \min_{s^i} \int_{{\bf
x}^i(s^i) = {\bf x}_1^i}^{{\bf x}_2^i} \sqrt{\delta {\bf x}^{i\ast T}
{\bf M} \delta {\bf x}^i} \ge 0$ to a given other trajectory ${\bf
x}_2^i$ in a metric ${\bf M(x}^i, i) = {\bf \Theta} ({\bf x}^i,
i)^{\ast T} \ {\bf \Theta}({\bf x}^i, i) \ge 0$ exponentially
converges to ${\bf x}_2^i$ within the bounds
\begin{equation}
\sigma_{\max} \ge \frac{s^{i+1}}{s^i} \ge \sigma_{\min}
\label{eq:exdisbound}
\end{equation}

$\sigma_{\max}$ ($\sigma_{\min}$) is defined as the largest (smallest)
singular value of the generalized Jacobian
$$
{\bf F}({\bf x}^i, i) \ = \ {\bf \Theta}^{(+1)} \frac{\partial
{\bf f}} {\partial {\bf x}^i} {\bf \Theta}^{-1}
$$
in the ball of radius $s^i$ around ${\bf x}_2^i$.

The system is said to be contracting (diverging) for uniformly
negative $\sigma_{\max} -1$ (uniformly positive $\sigma_{\min}
-1$). The system is said to be contracting (diverging) for negative
$\sigma_{\max} -1$ (positive $\sigma_{\min} -1$) and indifferent for
$\sigma_{\max} = \sigma_{\min} = 1$.

For a u.p.d. and bounded metric also the distance $\min_{s^i} \int_{{\bf
x}^i(s^i)= {\bf x}_1^i}^{{\bf x}_2^i} \sqrt{\delta {\bf x}^{i \ast T}
\delta {\bf x}^i}$ converges uniformly exponentially with the rates
above, where however initial overshoots can occure.

Note that the region of convergence of two arbitrary trajecctories
with distance dynamics $s^{i+1}/s^i$ in (\ref{eq:exdisbound}) can be
extended beyond the contracting region with Lyapunovs direct method
for the specific case that explicite orthonormal Cartesian coordinates
${\bf z}^i({\bf x}^i, i)$ with dimension $\ge N$ exist as
\begin{equation}
\frac{s^{i+1}}{s^i} = \frac{ \left({\bf z}_1^{i+1} - {\bf z}_2^{i+1}
\right)^{\ast T} \left({\bf z}_1^{i+1} - {\bf z}_2^{i+1} \right) } {
\left({\bf z}_1^i - {\bf z}_2^i \right)^{\ast T} \left({\bf z}_1^i -
  {\bf z}_2^i \right) }
\label{eq:si1si}
\end{equation}
\label{th:theoremFdis}
\end{theorem}

This theorem can be regarded as a time-varying, complex generalization
of the contraction mapping theorem (see e.g. \cite{Bert}) to a
general metric. This lead to the notation {\it Contraction Theory}.

\section{Continuous-time controllers} \label{contcontroller}

In this section we consider $\forall t \ge 0$ a smooth $n$-th order
real dynamic system in controllability form
$$ 
{\bf x}^{(n)} = {\bf f}({\bf x}, ..., {\bf x}^{(n-1)}, t) + {\bf
G}({\bf x}, ..., {\bf x}^{(n-1)}, t) {\bf u}
$$
with $N$-dimensional position ${\bf x}$, $M$-dimensional control input
${\bf u}$ and time $t$. The controllability conditions under which a
general continuous, non-linear, dynamic system can be transformed in
the form above is well established for feedback linearizable systems
(see e.g. \cite{flatness} or \cite{Reboulet}).

Let us now generalize the well-known LTI eigenvalue-placement in
Jordan form to the placement of the hierarchical complex dynamics  
\begin{equation}
\dot{\bf z}_j = \int {\bf \Lambda}_j ({\bf z}_j, t) d{\bf z}_j +
\Re({\bf z}_{j+1}) \ \mbox{for } j = 1, ..., p \label{eq:Jordancontroller}
\end{equation}
with $\Re({\bf z}_1) = {\bf x}, {\bf z}_{p+1} = 0$ and where $p$ is given
by $n$ minus the number of complex contraction rate matrices ${\bf
\Lambda}_j$.
Taking the variation of the above implies the time- or state-dependent
complex contraction rate matrices ${\bf \Lambda}_j ({\bf z}_j, t)$ in
\begin{equation}
\frac{d}{dt} \delta {\bf z}_j \ = \ {\bf \Lambda}_j ({\bf z}_j, t)
\delta {\bf z}_j + \Re(\delta {\bf z}_{j+1}) \ \mbox{for } j = 1, ...,
p \nonumber
\end{equation}
According to Theorem \ref {th:theoremF} is the stability of this
hierarchy given by the definiteness of the Hermitian part of ${\bf
\Lambda}_j$. 

Substituting the $p$ dynamics 
(\ref{eq:Jordancontroller}) recursively in each other leads to
\begin{theorem} 
Given the smooth $n$-th order dynamic system in controllability form
\begin{equation}
{\bf x}^{(n)} = {\bf f}({\bf x}, ..., {\bf x}^{(n-1)}, t) + {\bf
  G}({\bf x}, ..., {\bf x}^{(n-1)}, t) {\bf u}
\label{eq:contcontroldynamics}
\end{equation}
with $N$-dimensional position ${\bf x}$, $M$-dimensional control input
${\bf u}$ and time $t$. 

A controller ${\bf u}$ that places the
complex, integrable contraction rates ${\bf \Lambda}_j({\bf z}_j, t)$
in the characteristic equation
\begin{equation}
\left(\frac{d}{dt} - \int \Lambda_p d \right)\Re ...
\left(\frac{d}{dt} - \int \Lambda_1 d \right) {\bf x} = {\bf 0}
\label{eq:contcontrolchar}
\end{equation}
with $\Re({\bf z}_1) = {\bf x}$ and ${\bf z}_{j+1} = \dot{\bf z}_j -
\int \Lambda_j d {\bf z}_j$ implies global contraction behaviour with
${\bf \Lambda}_j({\bf z}_j, t)$ according to Theorem
\ref{th:theoremF}.

$p$ is here given by $n$ minus the number of complex contraction rate
matrices ${\bf \Lambda}_j$ and $\Re$ applies to its left-hand
term. The open integral $\int$ implies a time-varying integration
constant that can be chosen to shape a desired trajectory in the flow
field without affecting the contraction
behaviour. \label{th:contcontroller}
\end{theorem}

The generalization to standard feedback linearization methods (see
e.g. \cite{flatness} or \cite{Reboulet}) is that we can choose state-
or time-dependent contraction rates ${\bf \Lambda}_j({\bf z}_j, t)$ to
simplify ${\bf u}$, to handle only piece-wisely controllable systems
or simply to achieve state- or time-dependent system performance.

In contrast to well-known gain-scheduling techniques (see
e.g. \cite{Lawrence}), who also intend to achieve state-dependent
stability behaviour, we can analytically proof global contraction
behaviour with ${\bf \Lambda}_j({\bf z}_j, t)$. Analytic robustness
guarantees to modelling uncertainties are given in section
\ref{continuoushigher}.

Note that (\ref{eq:contcontrolchar}) can be modally solved as
$$
\delta {\bf z}_j(t) = \Phi(0,t) \int_o^t \Re(\delta {\bf
z}_{j+1}(\tau)) \Phi(\tau, 0) d \tau + \Phi(0,t) \delta {\bf z}_j^o
$$
with the transition matrix $\Phi(0,t)$ in equation
(\ref{eq:transition}) which can be analytically over/under-bounded
with Theorem \ref{th:theoremF}. This extends the well-established LTI
convolution principle to state- and time-dependent contraction rates.

Let us first consider LTV systems before we go to the non-linear case:
\Example{}{Consider the second-order real, time-varying dynamics
$$
\ddot{\bf x} + {\bf D}(t) \dot{\bf x} + {\bf K}(t) {\bf x} = {\bf
  u}(t) 
$$

Real contraction rates ${\bf \Lambda}_1(t)$ and ${\bf \Lambda}_2(t)$
imply with the characteristic equation (\ref{eq:contcontrolchar}) in
Theorem \ref{th:contcontroller}
\begin{eqnarray}
{\bf D}(t) &=& -{\bf \Lambda}_1 - {\bf \Lambda}_2 \nonumber \\
{\bf K}(t) &=& {\bf \Lambda}_2 {\bf \Lambda}_1 - \dot{\bf \Lambda}_1
\nonumber 
\end{eqnarray}

A complex contraction rate ${\bf \Lambda}_1(t)$ in
$$
\delta \dot{\bf z}_1 = {\bf \Lambda}_1 \delta {\bf z}_1 
$$
implies the real dynamics
\begin{eqnarray}
\delta \dot{\bf x} &=& {\bf Re} \delta {\bf x} + {\bf Im} \delta {\bf
  y} \nonumber \\
\delta \dot{\bf y} &=& -{\bf Im} \delta {\bf x} + {\bf Re} \delta {\bf
  y} \nonumber
\end{eqnarray}
with $\delta {\bf x} = \Re(\delta {\bf z}_1)$, $\delta {\bf y} =
\Im(\delta {\bf z}_1)$, ${\bf Re} = \Re({\bf \Lambda}_1)$ and 
${\bf Im} = \Im({\bf \Lambda}_1)$. Rewriting the above as second-order
dynamics in $\delta {\bf x}$ implies
\begin{eqnarray}
{\bf D}(t) &=& - 2 {\bf Re} - \dot{\bf Im}{\bf Im}^{-1} \nonumber \\
{\bf K}(t) &=& - {\bf Re} {\bf Re} + {\bf Im}{\bf Im} - {\bf D} {\bf
  Re}  - \dot{\bf Re} \nonumber
\end{eqnarray}
Note that only the additional time-derivative of ${\bf \Lambda}_1$
make this analytic stability result correct in comparison to a
standard LTI approximation of the LTV system.}{complexLTV}
Let us now consider real non-linear systems before we go to the complex
non-linear case:
\Example{}{Let us now schedule $\lambda_1(z_1, Ma, q_c)$ and
$\lambda_2(z_2, Ma, q_c)$ in example \Ex{accontroller} (see
e.g. \cite{Moritz}) in the characteristic equation
(\ref{eq:contcontrolchar}) in Theorem \ref{th:contcontroller}
$$
\left(\frac{d}{dt} - \int \lambda_2 d \right) \left(\dot{\alpha} - \int
\lambda_1 d \alpha \right) = 0
$$
with $z_1 = \alpha, z_2 = \dot{z}_1 - \int \lambda_1 dz_1$ which is
equivalent to
\begin{eqnarray}
\ddot{\alpha} &=& - D(q_c, Ma) \dot{\alpha} - K(q_c, Ma) \alpha +
G(q_c, Ma) u \nonumber \\
&=& \frac{d}{dt} \int \lambda_1 d z_1 + \int \lambda_2 d z_2 \nonumber
\end{eqnarray}
where the time-varying integration constant can be chosen to achieve
tracking-behaviour of the controller.

Again the difference to standard gain-scheduling techniques (see
e.g. \cite{Lawrence}) is the integration over $\lambda_2$ and the time
derivative of $\lambda_1$. Only with these terms exponential
convergence with the eigenvalues is given.}{extaccontroller}
Let us now go to complex state-dependent contraction rates. This
extension allows to achieve global stability for partially
controllable systems as e.g. the inverted pendulum.

\Example{}{Let us now place for the inverted pendulum without
gravity 
$$
\ddot{x} = u \cos(x)
$$
in figure \ref{fig:InvPend} the complex contraction rate $\lambda =
\cos(z)$ with $z = x + i y$ of the complex dynamics
$$
\dot{z} = \sin(z)
$$ 
We assume without loss of generality $- \pi \le x \le \pi$.
The first-order complex dynamics is equivalent to
\begin{eqnarray}
\dot{x} &=& \sin(x) \cosh(y) \nonumber \\
\dot{y} &=& \cos(x) \sinh(y) \nonumber
\end{eqnarray}
whose real second-order plant dynamics is 
$$
\ddot{x} = \cos(x) \dot{x} \cosh(y) + \sin(x) \sinh(y) \dot{y}
= \cos(x) \sin(x) cosh(2 y)
$$
with the control input $u = {\sin x} \cosh(2 y)$ that stays bounded for
bounded $y$. 

The chosen contraction rate $\lambda$ is according to Theorem
\ref{th:theoremF} diverging for the lower positions $\cos x > 0$ and
contracting for the upper positions $\cos x < 0$.

The convergence rate of an arbitrary trajectory $z_1$ to the lower
pendulum position $z_2=0$ is
\begin{equation}
\frac{\dot{s}}{s} = \frac{ \Re \left( \dot{z}_1^{\ast} z_1 \right)} {
  z_1^{\ast} z_1 } = \frac{\sin(x) x \cosh(y) + \cos(x) \sinh(y)
  y}{x^2 + y^2} \ge 0
\end{equation}
according to (\ref{eq:dss}) Theorem \ref{th:theoremF}. We can see that
the upper (lower) pendulum position is globally stable (unstable)
except the trajectory that starts exactly at the lower (upper)
pendulum position. The corresponding complex dynamics is illustrated
in figure \ref{fig:sin}. 
\begin{figure}
\begin{center}
\includegraphics[scale=0.4]{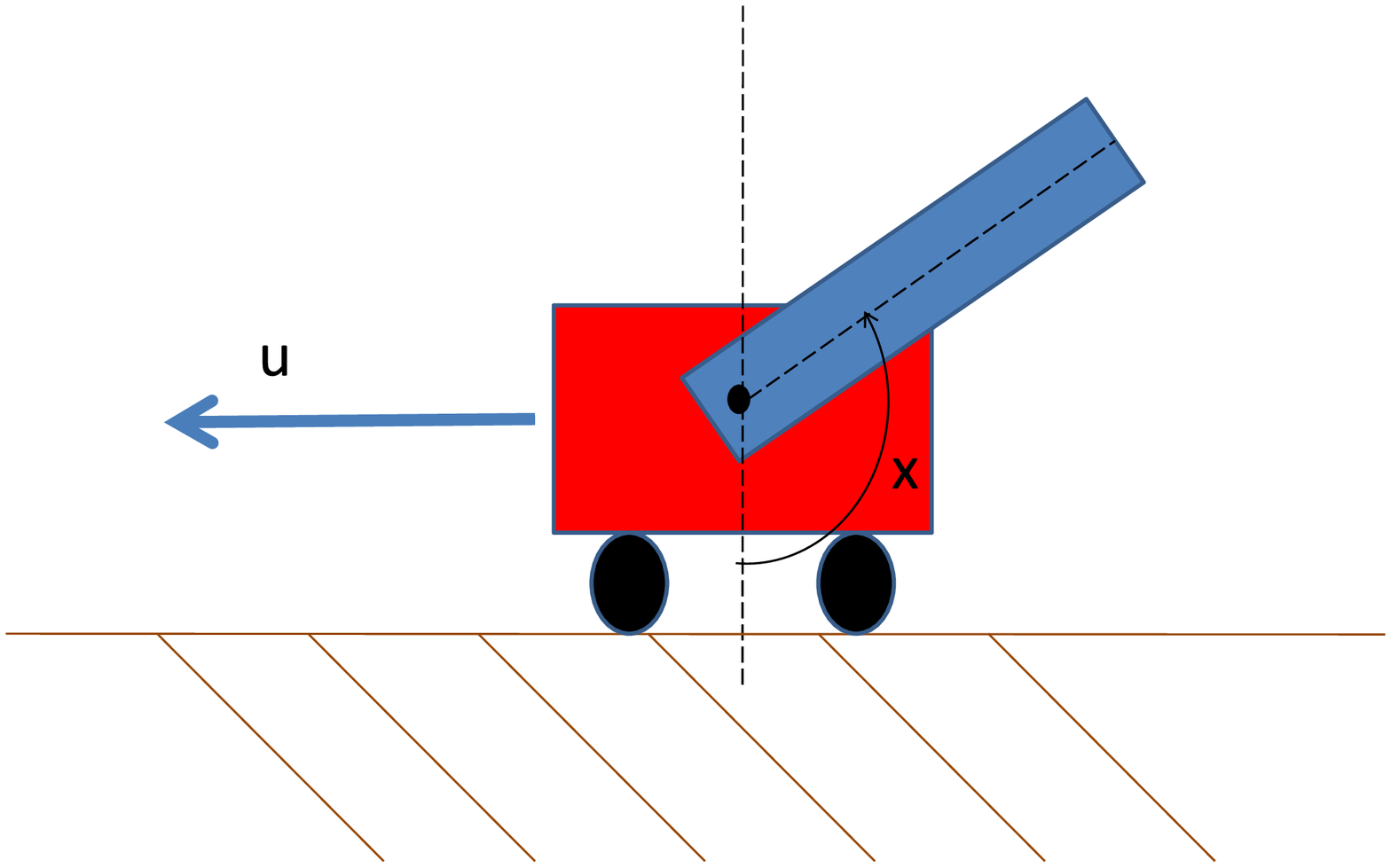}
\end{center}
\caption{Inverted pendulum}
\label{fig:InvPend}
\end{figure}

\begin{figure}
\begin{center}
\includegraphics[scale=0.4]{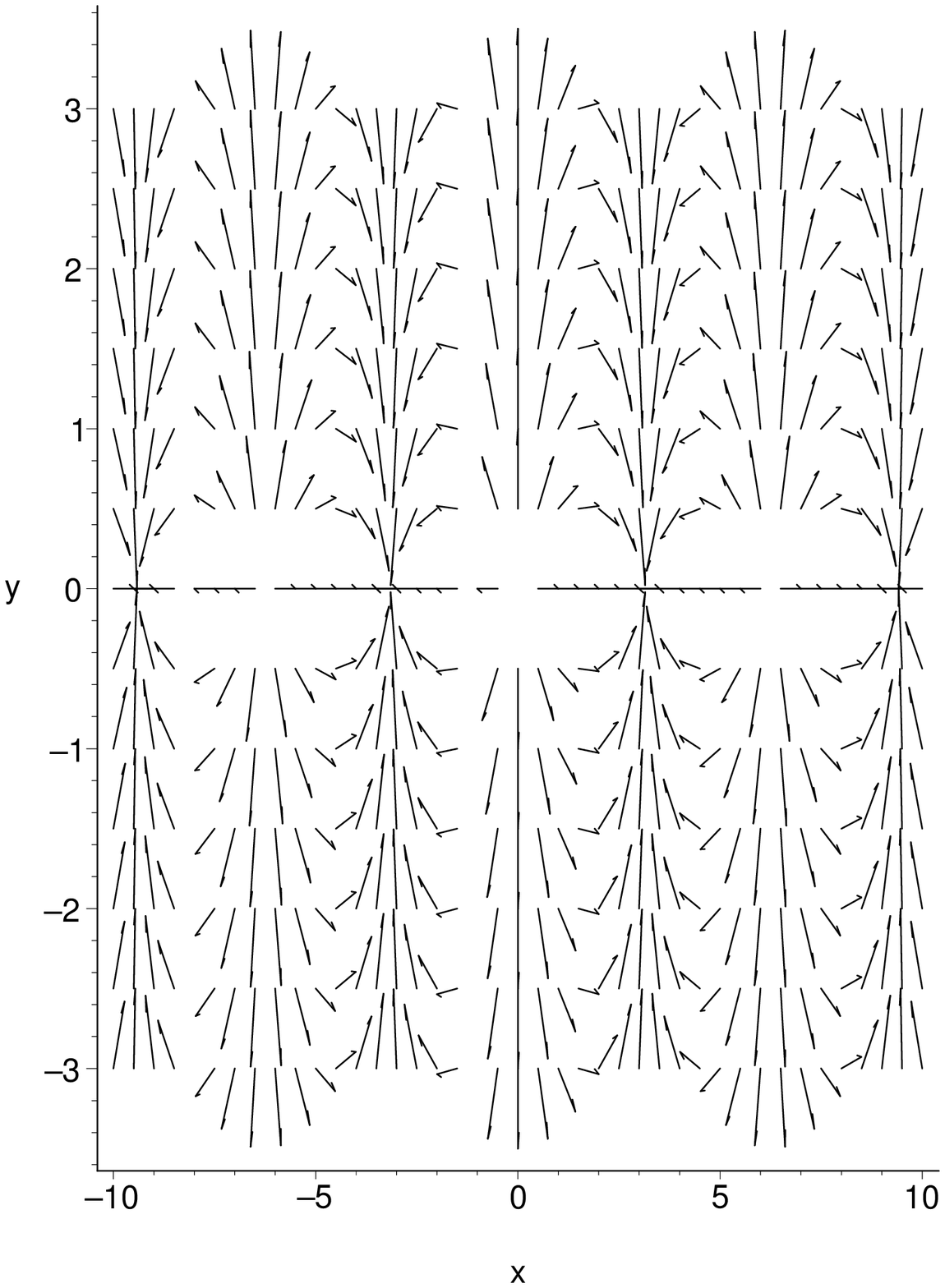}
\end{center}
\caption{Sinus complex state space dynamics of the inverted pendulum}
\label{fig:sin}
\end{figure}

Let us now choose alternatively the complex dynamics
$$
\dot{z} = (1+i) \sin(\frac{z}{2})
$$
with $\lambda_1 = \frac{1+i}{2} \cos(\frac{z}{2})$. The above is
equivalent to
\begin{eqnarray}
\dot{x} &=& \sin(\frac{x}{2}) \cosh(\frac{y}{2}) - \cos(\frac{x}{2})
\sinh(\frac{y}{2}) \nonumber \\
\dot{y} &=& \sin(\frac{x}{2}) \cosh(\frac{y}{2}) + \cos(\frac{x}{2})
\sinh(\frac{y}{2}) \nonumber
\end{eqnarray}
whose real second-order plant dynamics is 
$$
\ddot{x} = - \frac{1}{2} \cos(x) sinh(y)
$$
with the control input $u = - \frac{1}{2} \sinh(y)$ that stays bounded
for bounded $y$. The complex dynamics is illustrated in figure
\ref{fig:sinhalf}. We can see that - as designed - every second upper
position is globally stable / unstable.
\begin{figure}
\begin{center}
\includegraphics[scale=0.4]{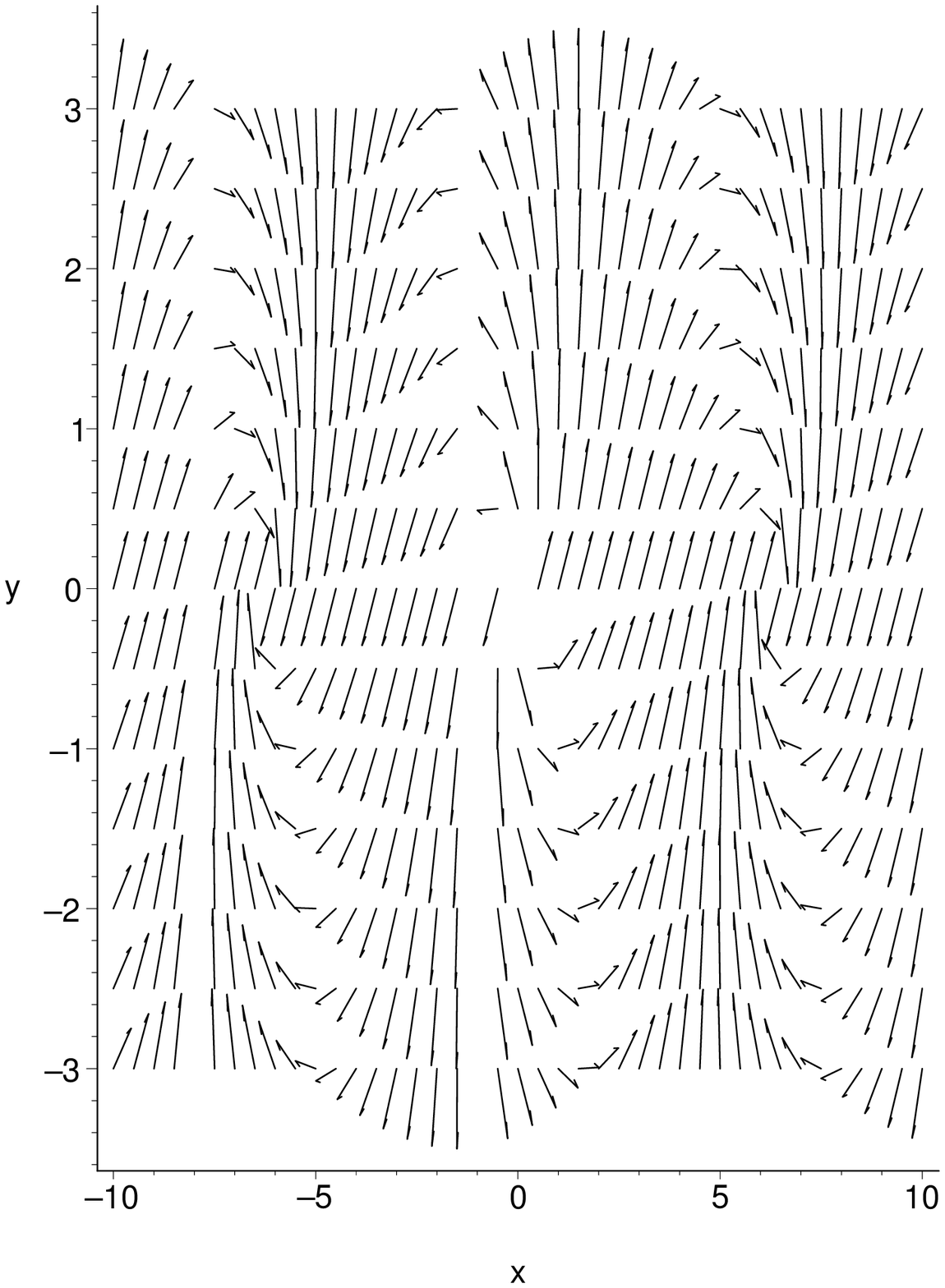}
\end{center}
\caption{Sinus half complex state space dynamics of the inverted
  pendulum}
\label{fig:sinhalf}
\end{figure}
}{complexsin}



\section{Continuous-time observers} \label{contobserver}

In this section we consider $\forall t \ge 0$ a smooth $n$-th order
dynamic system in observability form
$$
{\bf x}^{(n)} = {\bf a}_1^{(n-1)}({\bf x}, t) + {\bf
a}_2^{(n-2)}({\bf x}, t) + ...  + {\bf a}_n({\bf x}, t)
$$
with $M$-dimensional measurement ${\bf y}({\bf x}, t)$,
$N$-dimensional state ${\bf x}$, $N$-dimensional non-linear plant
dynamics ${\bf a}_j({\bf x}, t)$ and time $t$, which is equivalent to
\begin{equation} 
\dot{{\bf x}}_j = {\bf x}_{j+1} + {\bf
a}_j(\hat{\bf x}, t) \ \mbox{for } j = 1, ..., n \label{eq:plant}
\end{equation}
with ${\bf x}_1 = {\bf x}$ and ${\bf x}_{n+1} = 0$ 

Let us now introduce the observer
\begin{equation} 
\dot{\hat{{\bf x}}}_j = \hat{\bf x}_{j+1} + {\bf
a}_j(\hat{\bf x}, t) + {\bf e}_j(\hat{\bf y}, t) - {\bf
e}_j({\bf y}, t) \ \mbox{for } j = 1, ..., n \nonumber
\end{equation}
with $\hat{\bf x}_1 = \hat{\bf x}$ and $\hat{\bf x}_{n+1} = 0$ that
allows to extend the plant dynamics ${\bf a}_j$ with a chosable
measurement feedback ${\bf e}_j$ in the equivalent $n$-th order observer
dynamics
\begin{equation}
\hat{\bf x}^{(n)} = \sum_{j=1}^n \left( {\bf a}_j(\hat{\bf x}, t)
+ {\bf e}_j(\hat{\bf y}, t) - {\bf e}_j({\bf y}, t) \right)^{(n-j)}
\label{eq:conthighobs} 
\end{equation}

Let us now generalize the well-known LTI eigenvalue-placement in
Jordan form to the placement of the hierachical complex dynamics
\begin{equation}
\dot{\hat{{\bf z}}}_j = \int {\bf \Lambda}_j (\hat{\bf z}_j, t) d
\hat{\bf z}_j +  \Re(\hat{\bf z}_{j+1}) \ \mbox{for } j = 1, ..., p-1
\label{eq:descontobs} 
\end{equation}
with $\Re(\hat{\bf z}_1) = \hat{\bf x}, \hat{\bf z}_{p+1} = 0$ and
where $p$ is given by $n$ minus the number of complex contraction rate
matrices ${\bf \Lambda}_j$. Taking the variation of the above implies
the time- or state-dependent complex contraction rate matrices ${\bf
\Lambda}_j$ in
\begin{equation}
\frac{d}{dt} \delta \hat{\bf z}_j \ = \ {\bf \Lambda}_j (\hat{\bf
z}_j, t) \delta \hat{\bf z}_j + \Re(\delta \hat{\bf z}_{j+1}) \
\mbox{for } j = 1, ..., p \nonumber
\end{equation}
According to Theorem \ref{th:theoremF} is the stability of this
hierachy given by the definiteness of the Hermitian part of ${\bf
  \Lambda}_j$.

Substituting the $p$ dynamics (\ref{eq:descontobs}) recursively in
each other leads to
\begin{theorem} 
Given the smooth $n$-th order dynamic system in observability form
\begin{equation}
{\bf x}^{(n)} = {\bf a}_1^{(n-1)}({\bf x}, t) + {\bf a}_2^{(n-2)}({\bf
x}, t) + ...  + {\bf a}_n({\bf x}, t) \label{eq:contobsdynamics}
\end{equation}
with $M$-dimensional measurement ${\bf y}({\bf x}, t)$, $N$-dimensional
state ${\bf x}$, $N$-dimensional non-linear plant
dynamics ${\bf a}_j({\bf x}, t)$ and time $t$.

An observer 
\begin{equation} 
\dot{\hat{{\bf x}}}_j = \hat{\bf x}_{j+1} + \left( {\bf a}_j(\hat{\bf
x}, t) + {\bf e}_j(\hat{\bf y}, t) - {\bf e}_j({\bf y}, t) \right) \
\mbox{for } j = 1, ..., n \label{eq:contobserver}
\end{equation}
with $\hat{\bf x}_1 = \hat{\bf x}$ and $\hat{\bf x}_{n+1} = {\bf 0}$
allows to place with the measurement feedback terms ${\bf
e}_j$ the time- or state-dependent, integrable, complex contraction
rate matrices ${\bf \Lambda}_j(\hat{\bf z}_j, t)$ in the
characteristic equation
\begin{equation}
\left(\frac{d}{dt} - \int \Lambda_p d \right)\Re ...
\left(\frac{d}{dt} - \int \Lambda_1 d \right) \hat{\bf x} = {\bf 0}
\label{eq:contobserverchar}
\end{equation}
with $\Re(\hat{\bf z}_1) = \hat{\bf x}, \Re(\hat{\bf z}_{j+1}) =
{\dot{\hat{\bf z}}}_j - \int {\bf \Lambda}_j (\hat{\bf z}_j, t) d
\hat{\bf z}_j$ 

The definiteness of the Hermitian part of ${\bf \Lambda}_j(\hat{\bf
z}_j, t)$ implies global contraction behaviour of the observer state
with ${\bf \Lambda}_j(\hat{\bf z}_j, t)$ to the plant state according
to Theorem \ref{th:theoremF}.

$p$ is given by $n$ minus the number of complex contraction rate
matrices ${\bf \Lambda}_j$ and $\Re$ applies to its left-hand
term. \label{th:contobserver}
\end{theorem}

This theorem generalizes the extended LTV Luenberger observer design
of constant eigenvalues (see e.g. \cite{Lohm1}, \cite{Mracek} or
\cite{Zeitz}) to non-linear or state-dependent contraction rates for
non-linear, time-varying systems. It provides a systematic observer
design technique compared to existing contraction observer designs
(see e.g.  \cite{Rouchon,Nguyen,inertial})

Note that the global controller in Theorem \ref{th:contcontroller}
that uses the state estimates of the global observer in Theorem
\ref{th:contobserver} satisfies a separation principle. Indeed,
subtracting the plant dynamics (\ref{eq:plant}), eventually extended by
a control input ${\bf G}({\bf y}, t) {\bf u}(\hat{\bf x}, t)$, from
the observer dynamics (\ref{eq:contobserver}), that is extended by the
same control input ${\bf G}({\bf y}, t) {\bf u}(\hat{\bf x}, t)$,
leads with $\tilde{\bf x} = \hat{\bf x} - {\bf x}$ and the mid-point
theorem to
\begin{equation} 
\dot{\tilde{{\bf x}}}_j = \tilde{\bf x}_{j+1} + \frac{\partial \left(
{\bf a}_{n-j} + {\bf e}_{n-j} \right)}{\partial {\bf x}} ({\bf \xi},
t) \tilde{\bf x} \ \mbox{for } j = 1, ..., n \nonumber
\end{equation}
with $\tilde{\bf x}_{n+1} = 0$ and where ${\bf \xi}$ is one point
between ${\bf x}$ and $\hat{\bf x}$. We can see that the Jacobian of
the error-dynamics of the observer is unchanged. Since $\frac{\partial
{\bf u}}{\partial {\bf x}}$ in Theorem \ref{th:contcontroller} is
bounded the controller represents a hierarchical system
\cite{Lohm1}. As a result is the convergence rate of the controller
unchanged as well.

Let us now show how a general $N n$ dimensional plant
$$
\dot{\underline{\bf x}} = {\bf f}(\underline{\bf x}, t)
$$
with $N$-dimensional measurement ${\bf y} = {\bf x}(\underline{\bf x},
t)$ can be transformed to the higher-order observability form
(\ref{eq:contobsdynamics}). A necessary condition is that the mapping
$$
\left( \begin{array}{c} {\bf y}(\underline{\bf x}, t) \\ \vdots \\
{\bf y}^{(n-1)}(\underline{\bf x}, t) \end{array} \right) 
$$
can be inverted to $\underline{\bf x}({\bf y}, ..., {\bf y}^{(n-1)},
t)$ such that we get an explicit dynamics (\ref{eq:contobsdynamics})
$$
{\bf x}^{(n)} = {\bf x}^{(n)}(\underline{\bf x}(({\bf x}, ..., {\bf
  x}^{(n-1)}, t), t)
$$
Hence a necessary (but not sufficient) observability condition is that
the observability matrix 
$$
{\bf O} = \left( \begin{array}{c} L^o {\bf c} \\ \vdots \\
L^{n-1} {\bf c} \end{array} \right)  
$$
with the Lie derivatives \cite{Lovelock} $L^o {\bf c} = \frac{\partial
{\bf y}}{\partial \underline{\bf x}}$ and $L^{j+1} {\bf c} =
\frac{\partial {\bf y}^{(j+1)}}{\partial \underline{\bf x}} =L^j {\bf
c} \frac{\partial {\bf f}}{\partial \underline{\bf x}} + \frac{d}{dt}
L^j {\bf c}$ has piece-wisely full rank. Note that for LTV systems it
is also sufficient.

Let us first consider a linear observer design with time-varying
contraction rates.
\Example{}{Consider the vertical channel dynamics of a navigation
  system
$$
\ddot{x} = a(t)
$$
with measured altitude $y=x$ and measured vertical acceleration $a(t)$.
We want to schedule with Theorem \ref{th:contobserver} the complex
eigenvalues $\lambda_1(Ma)$ and $\lambda_2(Ma)$ with $Ma(t)$ in the
observer (\ref{eq:contobserver})
$$
\left( \begin{array}{c} \dot{\hat{x}}_1 \\ \dot{\hat{x}}_2 \end{array}
\right) = \left( \begin{array}{c} \hat{x}_2 \\ a(t) \end{array}
\right) + \left( \begin{array}{c} e_1(Ma) \\ e_2(Ma) \end{array}
\right) \left(\hat{y} - y \right)
$$
with $\hat{x} = \hat{x}_1$ to optimize the vertical channel
performance for changing altitude measurement accuracy in sub-, trans-
and supersonic. Comparing the second-order observer error dynamics 
$$
\ddot{\tilde{x}} = \frac{d}{dt} \left( e_1(Ma) \tilde{x}
\right) + e_2(Ma) \tilde{x} 
$$
with $\tilde{x} = \hat{x} - x$ to the characteristic equation
(\ref{eq:contobserverchar})
\begin{equation}
\left(\frac{d}{dt} - \lambda_2 \right) \left(\frac{d}{dt} - \lambda_1
\right) \tilde{x} = \ddot{\tilde{x}} - \frac{d}{dt} \left( ( \lambda_1
+ \lambda_2) \tilde{x} \right) + ( \lambda_2 \lambda_1 +
\dot{\lambda}_2) \tilde{x} = 0 \nonumber
\end{equation}
leads to
\begin{eqnarray}
e_1(Ma) &=& \lambda_1 + \lambda_2 \nonumber \\
e_2(Ma, \dot{Ma}) &=& - \lambda_2 \lambda_1 - \dot{\lambda}_2
\nonumber  
\end{eqnarray}

The difference to standard gain-scheduling techniques (see
e.g. \cite{Lawrence}) is the term $\dot{\lambda}_2$ in the feedback
gain computation. Only with this term exponential convergence with the
eigenvalues $\lambda_1$ and $\lambda_2$ is given.}{verticalchannel}
Let us consider now observer designs for non-linear systems with
time-varying contraction rates:
\Example{}{Consider the temperature-dependent reaction $A \rightarrow
B$ in a closed tank
$$
\frac{d}{dt}
\left( 
\begin{array}{c}
c_A \\
T 
\end{array}
\right) =
\left( 
\begin{array}{c}
-1 \\
-10
\end{array}
\right)
e^{-\frac{E}{T}} c_A
$$
with $c_A$ the concentration of A, $y = T$ the measured temperature,
and $E$ the specific activation energy, where we want to build an
observer with designed contraction rates $\lambda_1(t), \lambda_2(t) <
0$.

This reaction dynamics is equivalent to the following second-order
dynamics in temperature
$$
\ddot{T} + \frac{-E}{T^2} \dot{T}^2 = -e^{-\frac{E}{T}} \dot{T} 
$$
Letting $x = \int_o^T e^{\frac{-E}{T}} dT$ yields the plant in
observability form 
$$
\ddot{x} = \dot{a}_1(x) 
$$
with $a_1(x) = - \int e^{-\frac{E}{\hat{T}(x)}} d x$. Let us design the
observer (\ref{eq:contobserver}) with estimate $\hat{x} = \hat{x}_1$
$$
\left( \begin{array}{c} \dot{\hat{x}}_1 \\ \dot{\hat{x}}_2 \end{array}
\right) = 
\left( \begin{array}{c} 
\hat{x}_2 + a_1  \\ 
0 \end{array} \right)
+ \left( \begin{array}{c} 
e_1(\hat{y}) - e_1(y) \\ e_2(\hat{y}) - e_2(y)
\end{array} \right)
$$
with designed time-varying contraction rates $\lambda_1(t),
\lambda_2(t) < 0$. Comparing the equivalent second-order observer
dynamics in $\hat{x}$
$$
\ddot{x} - \ddot{\hat{x}} = \frac{d}{dt} \left( a_1(\hat{x}) -a_1(x) +
e_1(\hat{x}_1) - e_1(x) \right) + e_2(\hat{x}_1) - e_2(x)
$$
with the characteristic equation (\ref{eq:contobserverchar})
\begin{equation}
\left(\frac{d}{dt} - \lambda_2 \right)
\left(\frac{d}{dt} - \lambda_1 \right) (\hat{x} - x)=  0 \nonumber
\end{equation}
leads to the non-linear feedback gains
\begin{eqnarray}
e_1(\hat{y}) + a_1(\hat{y}) &=& (\lambda_1 + \lambda_2) \hat{y}
\nonumber \\
e_2(\hat{y}) &=& -( \lambda_1 \lambda_2 + \dot{\lambda}_2 )
\hat{y} \nonumber 
\end{eqnarray}
}{chemical}
The following example gives an explicit equation for the feedback
gains of time-dependent contraction rates:
\Example{}{Consider the $n$-dimensional non-linear system dynamics
$$
{\bf x}^{(n)} = {\bf a}_1^{(n-1)}({\bf x}, t) + {\bf a}_2^{(n-2)}({\bf
  x}, t) + \ldots + {\bf a}_n({\bf x}, t) 
$$
with non-linear plant dynamics ${\bf a}_j({\bf x}, t)$ and measurement
vector ${\bf y}({\bf x},t)$. 

Comparing the $n$-th order dynamics (\ref{eq:conthighobs}) of the
observer (\ref{eq:contobserver}) to the characteristic equation
(\ref{eq:contobserverchar}) of real time-varying contraction rates
${\bf \Lambda}_j(t)$ implies the feedback gains
\begin{eqnarray}
{\bf e}_1(\hat{\bf y}, t) + {\bf a}_1(\hat{\bf x}, t) &=&
\sum_{j=1}^n {\bf \Lambda}_j \hat{\bf x} \nonumber \\ &\vdots&
\nonumber \\
{\bf e}_n(\hat{\bf y}, t) + {\bf a}_n(\hat{\bf x}, t) &=& 
ad_{{\bf \Lambda}_n} ... (ad_{{\bf \Lambda}_2} {\bf \Lambda}_1)
\hat{\bf x} \nonumber 
\end{eqnarray}
with $ad_{\bf G} {\bf F} = -{\bf G} {\bf F} - \dot{\bf G}$.
}{pseudolinearobserver}
Finally let us consider a non-linear observer with state-dependent
contraction rates:
\Example{}{Consider the Van-der-Pol oscillator
$$
\ddot{x} = \dot{a}_1(x) + a_2(x,t)
$$
with $a_1(x) = x - \frac{x^3}{3}$ and measured $y=x$. We want to build
an observer (\ref{eq:contobserver}) with estimate $\hat{x} = \hat{x}_1$
$$
\left( \begin{array}{c} \dot{\hat{x}}_1 \\ \dot{\hat{x}}_2 \end{array}
\right) = \left( \begin{array}{c} \hat{x}_2 + a_1(\hat{x}_1) \\ 
a_2(\hat{x}_1,t) \end{array} \right) + 
\left( \begin{array}{c} e_1(\hat{y}) - e_1(y) \\
e_2(\hat{y}) - e_2(y) \end{array} \right)
$$
with designed contraction rates $\lambda_1(\hat{x}), \lambda_2 <
0$. Comparing the equivalent second-order observer dynamics in
$\hat{x}$
$$
\ddot{x} - \ddot{\hat{x}} = \frac{d}{dt} \left( a_1(\hat{x}) -a_1(x) +
e_1(\hat{x}_1) - e_1(x) \right) + a_2(x,t) - a_2(x,t) + e_2(\hat{x}_1)
- e_2(x)
$$
with the characteristic equation (\ref{eq:contobserverchar})
\begin{eqnarray}
\left(\frac{d}{dt} - \int \lambda_2 d \right)
\left(\frac{d}{dt} - \int \lambda_1 d \right) \hat{x} &=& \nonumber \\
\ddot{x} - \ddot{\hat{x}} - \frac{d}{dt} \int_x^{\hat{x}} (\lambda_1 +
\lambda_2) d \hat{x} + \lambda_2 \int_x^{\hat{x}} \lambda_1 d
\hat{x} &=& 0 \nonumber
\end{eqnarray}
leads to the non-linear feedback gains
\begin{eqnarray}
e_1(\hat{y}) + a_1(\hat{y}) &=& \int (\lambda_1 + \lambda_2) d \hat{y}
\nonumber \\
e_2(\hat{y},t) + a_2(\hat{y},t) &=& - \lambda_2 \int \lambda_1 d
\hat{y} \nonumber
\end{eqnarray}
}{VanderPol}

\section{Continuous higher-order analysis and robustness}
\label{continuoushigher}

Consider for $t \ge 0$ the $n$-dimensional ($n \ge 1$) system
$$
{\bf x}^{(n)} = {\bf f}({\bf x}, ..., {\bf x}^{(n-1)}, t)
$$
with $N$-dimensional position ${\bf x}$. 

In Theorem \ref{th:contcontroller} and \ref{th:contobserver} the
characteristic equation of the dynamics above is zero since we use the
observer or controller feedback to precisely achieve the
characteristic equation. For such a given controller or observer an
additional modelling uncertainty ${\bf d}$ may have to be considered
on top to the designed characteristic dynamics. This introduces the
idea of the existence of a perturbation ${\bf d}$ in the
characteristic equation if we analyse a given ODE.

Based on this thought let us approximate this
dynamics with the complex, integrable contraction rates ${\bf
\Lambda}_j({\bf z}_j, t)$ - that eventually correspond to the designed
contraction rates - in the distorted characteristic equation
\begin{equation}
\left(\frac{d}{dt} - \int \Lambda_p d \right)\Re ...
\left(\frac{d}{dt} - \int \Lambda_1 d \right) {\bf x} = {\bf d}({\bf
  z}_p, .. {\bf z}_1, t) \nonumber
\end{equation}
with $\Re({\bf z}_1) = {\bf x}$ and $\Re({\bf z}_{j+1}) = \dot{\bf
z}_j - \int \Lambda_j d {\bf z}_j$.

Taking the variation of the above we get
\begin{equation}
\delta \dot{\bf z}_p - {\bf \Lambda}_p \delta {\bf z}_p =
\frac{\partial {\bf d}}{\partial {\bf z}_p} \delta {\bf z}_p + \ ... \
+ \frac{\partial {\bf d}}{\partial {\bf z}_1} \delta {\bf z}_1
\nonumber
\end{equation}
The main idea is to construct in the following an exponential
bound on the virtual displacement $\delta {\bf z}_1$ over $p$
time-derivatives, rather than over the first time-derivative as in
\cite{Lohm1}.

Let us first bound the higher-order term part by taking the norm of
the above
\begin{equation}
\left|\delta \dot{\bf z}_p - {\bf \Lambda}_p \delta {\bf z}_p \right|
\le \left|\frac{\partial {\bf d}}{\partial {\bf z}_p}\right| |\delta
{\bf z}_p| + \ldots + \left|\frac{\partial {\bf d}}{\partial {\bf
z}_1}\right| |\delta {\bf z}_1| \label{eq:normdyn}
\end{equation}
where now and in the following the norm of a matrix is the largest
singular value of that matrix and the norm of a vector is the
root of the vector multiplied with its conjungate vector.  

Let us now select a real $\eta({\bf z}_1(t),...{\bf z}_p(t), t)$ that
fulfils
\begin{equation} 
L^p \eta \ge |\frac{\partial {\bf d}}{\partial {\bf z}_p}| L^p
\eta + \ \ldots \ + |\frac{\partial {\bf d}}{\partial {\bf z}_1}|
\label{eq:characteristic}
\end{equation}
$\forall t \ge 0$ with $L^{j+1} \eta = \dot{L^j \eta} + \eta L^j \eta
\ge 0, L^o \eta = 1$ and let us bound the initial conditions at $t=0$
with real and constant $K \ge 0$ as
\begin{equation} 
|\delta {\bf z}_j| \le K \ L^{j-1} \eta \ e^{\int_0^t (\eta +
\lambda_{max})
  dt}, \ 1 \le j \le p \label{eq:dxn}
\end{equation}
where $\lambda_{max}({\bf z}_1(t),...{\bf z}_p(t), t)$ is the largest
eigenvalue of the Hermitian part of all ${\bf \Lambda}_j({\bf z}_j,
t)$. Hence with (\ref{eq:characteristic}) and (\ref{eq:dxn}) we can
bound (\ref{eq:normdyn}) at $t=0$ as
\begin{eqnarray}
|\delta \dot{\bf z}_p - {\bf \Lambda}_p \delta {\bf z}_p| &\le& K \
e^{\int_0^t (\eta + \lambda_{max}) dt} \left( |\frac{\partial {\bf
d}}{\partial {\bf z}_p}| L^p \eta + \ \ldots \ +
|\frac{\partial {\bf d}}{\partial {\bf z}_1}| \right) \nonumber \\
&\le& K \ L^p \eta \ e^{\int_0^t (\eta + \lambda_{max})}
\label{eq:dxn1}
\end{eqnarray}
Theorem \ref{th:theoremF} on $\delta \dot{\bf z}_j = {\bf \Lambda}_j
\delta {\bf z}_j + \Re( \delta {\bf z}_{j+1} )$ with the bounded
distortion (\ref{eq:dxn}) and (\ref{eq:dxn1}) implies at $t=0$
\begin{eqnarray}
|\delta {\bf z}_j(t+dt)| &\le& |\delta {\bf z}_j(t)| + K \ (
\lambda_{max} L^{j-1} \eta(t) + L^j \eta(t)) \ e^{\int_0^t (\eta +
\lambda_{max})} dt \nonumber \\ &\le& K \ L^{j-1} \eta(t+dt) \
e^{\int_0^{t+dt} (\eta + \lambda_{max})}, 1 \le j \le p \nonumber
\end{eqnarray}
which implies with complete induction that (\ref{eq:dxn}) and
(\ref{eq:dxn1}) hold $\forall t \ge 0$. Using the above this allow to
conclude:
\begin{theorem}
Consider for $t \ge 0$ the $n$-dimensional ($n \ge 1$) system
$$
{\bf x}^{(n)} = {\bf f}({\bf x}, ..., {\bf x}^{(n-1)}, t)
$$
with $N$-dimensional position ${\bf x}$. 

Let us approximate the above dynamics with the integrable complex
contraction rates ${\bf \Lambda}_j({\bf z}_j, t)$ in
the distorted characteristic equation
\begin{equation}
\left(\frac{d}{dt} - \int \Lambda_p d \right)\Re ...
\left(\frac{d}{dt} - \int \Lambda_1 d \right) {\bf x} = {\bf d}({\bf
  z}_p, .. {\bf z}_1, t) \
\label{eq:contchardist}
\end{equation}
with $\Re({\bf z}_1) = {\bf x}$ and ${\bf z}_{j+1} = \dot{\bf z}_j -
\int \Lambda_j d {\bf z}_j$. 

Bounding the effect of the distortion ${\bf d}$ with a real $\eta({\bf
  z}_1(t),...{\bf z}_p(t), t)$ that fulfils $\forall t \ge 0$
\begin{equation} 
L^p \eta^p \ge |\frac{\partial {\bf d}}{\partial {\bf z}_p}| L^p
\eta + \ \ldots \ + |\frac{\partial {\bf d}}{\partial {\bf z}_1}|
\label{eq:theoremcharacteristic}
\end{equation}
with $L^{j+1} \eta = \dot{L^j \eta} + \eta L^j \eta
\ge 0, L^o \eta = 1$ and the largest eigenvalue $\lambda_{max}({\bf
z}_1(t),...{\bf z}_p(t), t)$ of the Hermitian part of all ${\bf
\Lambda}_j({\bf z}_j, t)$ leads to global contraction behaviour with
\begin{equation}
{\bf \Lambda}_j({\bf z}_j, t) + \eta \label{eq:lambdaeta}
\end{equation}
according to Theorem \ref{th:theoremF} where initial overshoots are
bounded $\forall t \ge 0$ with constant $K > 0$ by
\begin{equation}  
|\delta {\bf z}_j | \le K \ L^{j-1} \eta(t=0) \ \ e^{\int_0^t
\eta + \lambda_{\max j} dt}, \ 1 \le j \le p \label{eq:definitionofK} 
\end{equation}

$p$ is given by $n$ minus the number of complex contraction rate
matrices ${\bf \Lambda}_j$ and $\Re$ applies to its left-hand
term.
\label{th:higherordercontinuous}
\end{theorem}

Interpreting ${\bf \Lambda}_j({\bf z}_j, t)$ as the desired
contraction rates in Theorem \ref{th:contcontroller} or
\ref{th:contobserver} allows to bound the potential instabilities
which come from modelling uncertainties of the plant.  I.e. it allows
to prove robustness for modelling uncertainties with a bounded
de-stabilizing divergence rate $\eta$.  Note that additional
time-varying errors in the control input do not affect the contraction
behaviour, but the desired trajectory ${\bf x}_d(t)$.

If we cannot design ${\bf \Lambda}_j({\bf z}_j, t)$ per feedback then
we have to approximate ${\bf \Lambda}_j$ to minimize the distortion
${\bf d}$, e.g. by transforming the higher-order system in its reduced
form (i.e. a form in which ${\bf f}$ is independent of ${\bf
x}^{(n-1)}$). This is illustrated in the following examples:
\Example{}{Consider the general (a)periodic dynamics
$$
\ddot{\bf x} + \frac{\partial \dot{U}}{\partial {\bf x}} +
\frac{\partial^2 U}{\partial {\bf x}^2}_{\min} \dot{\bf x} +
\frac{\partial V}{\partial {\bf x}} = 0
$$
with potentials $U({\bf x}, t), V({\bf x}, t)$, $N$-dimensional
position ${\bf x}$ and where we assume without los of generality
$\frac{\partial^2 U}{\partial {\bf x}^2}_{\min}(t) = \min \left(
\frac{\partial^2 U}{\partial {\bf x}^2} \right)$ $\forall {\bf x}$ at
a given $t$. The virtual dynamics is
$$
\delta \ddot{\bf x} + \left( \frac{\partial^2 U}{\partial {\bf x}^2} +
\frac{\partial^2 U}{\partial {\bf x}^2}_{\min} \right) \delta \dot{\bf
x} + \left( \frac{\partial^2 V}{\partial {\bf x}^2} + \frac{\partial^2
\dot{U}}{\partial {\bf x}^2} \right) \delta {\bf x} = 0
$$
The distorted characteristic equation (\ref{eq:contchardist}) is with
${\bf \Lambda}_1 = - \frac{\partial^2 U}{\partial {\bf x}^2},
{\bf \Lambda}_2 = -\frac{\partial^2 U}{\partial {\bf x}^2}_{\min}$
\begin{equation}
\left(\frac{d}{dt} - {\bf \Lambda}_2 \right) \left(\frac{d}{dt} - {\bf
\Lambda}_1 \right) \delta {\bf x} = \left( \frac{\partial^2
U}{\partial {\bf x}^2}_{\min} \frac{\partial^2 U}{\partial {\bf x}^2}
- \frac{\partial^2 V}{\partial {\bf x}^2} \right) \delta {\bf x}
\nonumber
\end{equation}

The remaining instability $\eta$ in (\ref{eq:theoremcharacteristic})
is then given 
$$
\dot{\eta} + \eta^2 \ge \left| \frac{\partial^2 U}{\partial {\bf
x}^2}_{\min} \frac{\partial^2 U}{\partial {\bf x}^2} -
\frac{\partial^2 V}{\partial {\bf x}^2} \right|
$$ 
We can hence bound with Theorem \ref{th:higherordercontinuous} the
contraction rates with 
$$
\frac{\partial^2 U}{\partial x^2}_{min} + \eta \le 
\frac{\partial^2 U}{\partial x^2}_{min} + \sqrt{ \left|
\frac{\partial^2 U}{\partial {\bf x}^2}_{\min} \frac{\partial^2
U}{\partial {\bf x}^2} - \frac{\partial^2 V}{\partial {\bf x}^2}
\right| }_{max}
$$
Note that for the scalar case with constant damping this condition is
equivalent to require that the complex poles $\frac{\partial^2
U}{\partial x^2} \pm \sqrt{ \frac{\partial^2 U}{\partial x^2}^2 -
\frac{\partial^2 V}{\partial x^2}(x,t) }$ lie $\forall x, t$ within
the $\pm 45^o$ quadrant of the left-half complex
plane.}{nonlinearsecondorder}

\Example{}{Consider the second-order non-linear system
$$
\ddot{x} + 2 \dot{x} + \frac{\partial V}{\partial x} = u(t)
$$
with potential energy $V = \ln \cosh (x - x_d(t))$, that increases the
stabilizing force $-\frac{\partial V}{\partial x} = -\tanh (x -
x_d(t))$ with the distance to the desired position $x_d(t)$. The
corresponding variational dynamics is
$$
\delta \ddot{x} + 2 \delta \dot{x} + \frac{1}{\cosh^2 (x - x_d(t))} \
\delta x = 0
$$
Since the LTI poles lie within $\pm 45°$ quadrant of the left-half
complex plane we can conclude with \Ex{nonlinearsecondorder} on
contraction behaviour.}{saturationfunction}



\section{Discrete-time controllers} \label{discontroller} 

In this section we consider $\forall i \ge 0$ a smooth $n$-th order
discrete system in controllability form
$$ 
{\bf x}^{i+n} = {\bf f}({\bf x}^i, ..., {\bf x}^{i+n-1}, i) + {\bf
G}({\bf x}^i, ..., {\bf x}^{i+n-1}, i) {\bf u}^i
$$
with $N$-dimensional position ${\bf x}^i$, $M$-dimensional control
input ${\bf u}^i$ and time index $i$. The controllability conditions
under which a general discrete, non-linear, dynamic system can be
transformed in the form above is well established for feedback
linearizable systems (see e.g. \cite{Lee, Nijmeijer}).

Let us now generalize the well-known LTI eigenvalue-placement in
Jordan form to the placement of the hierarchical complex dynamics
\begin{equation}
{\bf z}_j^{i+1} = \int {\bf \Sigma}_j ({\bf z}_j^i, i) d{\bf z}_j^i +
\Re({\bf z}_{j+1}^i) \ \mbox{for } j = 1, ..., p \label{eq:disclam}
\end{equation}
with $\Re({\bf z}_1^i) = {\bf x}^i, {\bf z}_{p+1}^i = 0$ and where $p$
is given by $n$ minus the number of complex contraction rate matrices
${\bf \Sigma}_j$. Taking the variation of the above implies the 
 time- or state-dependent complex contraction rate matrices ${\bf
   \Sigma}_j$ in 
\begin{equation}
\delta {\bf z}_j^{i+1} \ = \ {\bf \Sigma}_j({\bf z}_j^i, i)
\delta {\bf z}_j^i + \Re(\delta {\bf z}_{j+1}^i) \nonumber 
\end{equation}
According to Theorem \ref {th:theoremFdis} is the
stability of this hierachy given by the singular values of of ${\bf
  \Sigma}_j$. 

Substituting the $p$ dynamics (\ref{eq:disclam}) recursively in each
other leads to:
\begin{theorem} 
Given the smooth $n$-th order system in controllability form
\begin{equation} 
{\bf x}^{i+n} = {\bf f}({\bf x}^i, ..., {\bf x}^{i+n-1}, i) + {\bf
G}({\bf x}^i, ..., {\bf x}^{i+n-1}, i) {\bf u}^i
\label{eq:discretesystem}
\end{equation}
with $N$-dimensional position ${\bf x}^i$, $M$-dimensional control
input ${\bf u}^i$ and time index $i$. 

A controller ${\bf u}^i$ that places the complex, integrable
contraction rates ${\bf \Sigma}_j({\bf z}_j^i, i)$ in the
characteristic equation
\begin{equation}
\left( ^{(+1)} - \int {\bf \Sigma}_p d \right)\Re ...
\left( ^{(+1)} - \int {\bf \Sigma}_1 d \right) {\bf x}^i = {\bf 0}
\label{eq:discontrolchar}
\end{equation}
with $\Re({\bf z}_1^i) = {\bf x}^i, \Re({\bf z}_{j+1}^i) = {\bf
z}_j^{i+1} - \int {\bf \Sigma}_j ({\bf z}_j^i, i) d{\bf z}_j^i$
implies global contraction behaviour with ${\bf \Sigma}_j({\bf z}_j^i,
i)$ according to Theorem \ref{th:theoremFdis}.

$p$ is here given by $n$ minus the number of complex contraction rate
matrices ${\bf \Sigma}_j$ and $\Re$ applies to its left-hand
term. The open integral $\int$ implies a time-varying integration
constant that can be chosen to shape a desired trajectory in the flow
field without affecting the contraction
behaviour. \label{th:discontroller}
\end{theorem}

The generalization to standard feedback linearization methods (see
e.g. \cite{Lee, Nijmeijer}) is that we can choose state- or
time-dependent contraction rates ${\bf \Sigma}_j({\bf z}_j^i, i)$ to
simplify ${\bf u}^i$, to handle only piece-wisely controllable systems
or simply to achieve a state- or time-dependent system performance.

In contrast to well-known gain-scheduling techniques (see
e.g. \cite{Lawrence}), who also intend to achieve state-dependent
stability behaviour, we can analytically proof global contraction
behaviour with ${\bf \Sigma}_j({\bf z}_j^i, i)$. Analytic robustness
guarantees to modelling uncertainties are given in section
\ref{discretehigher}.

Note that (\ref{eq:discontrolchar}) can be modally solved as
$$
\delta {\bf z}_j^{i+1} = \sum_{l=0}^i \left( \Phi(l+1, i) \Re(
\delta {\bf z}_{j+1}^l) \right) + \Phi(0,i) \delta {\bf z}_j^o
$$
with the transition matrix $\Phi(0,i)$ in equation
(\ref{eq:disctransition}) which can be analytically over/under-bounded
with Theorem \ref{th:theoremFdis}. This extends the well-established
LTI convolution principle to state- and time-dependent contraction
rates.

Let us first consider LTV systems before we go to real and then
complex non-linear systems:
\Example{}{Consider the second-order real, time-varying dynamics
$$
{\bf x}^{i+2} + {\bf D}(i) {\bf x}^{i+1} + {\bf K}(i) {\bf x}^i =
{\bf u}(i)
$$
Real contraction rates ${\bf \Sigma}_1(i)$ and ${\bf \Sigma}_2(i)$
imply with the characteristic equation (\ref{eq:discontrolchar}) in
Theorem \ref{th:discontroller}
\begin{eqnarray}
{\bf D}(i) &=& -{\bf \Sigma}_1^{(+1)} - {\bf \Sigma}_2 \nonumber \\
{\bf K}(i) &=& {\bf \Sigma}_2 {\bf \Sigma}_1 \nonumber 
\end{eqnarray}

A complex contraction rate ${\bf \Sigma}_1(i)$ in
$$
\delta {\bf z}_1^{i+1} = {\bf \Sigma}_1(i) \delta {\bf z}_1^i 
$$
implies the real dynamics
\begin{eqnarray}
\delta {\bf x}^{i+1} &=& {\bf Re}^i \delta {\bf x}^i + {\bf Im}^i
\delta {\bf y}^i \nonumber \\
\delta {\bf y}^{i+1} &=& -{\bf Im}^i \delta {\bf x}^i + {\bf Re}^i
\delta {\bf y}^i \nonumber
\end{eqnarray}
with $\delta {\bf x}^i = \Re(\delta {\bf z}_1^i)$, $\delta {\bf y}^i =
\Im(\delta {\bf z}_1^i)$, ${\bf Re} = \Re({\bf \Sigma}_1)$ and 
${\bf Im} = \Im({\bf \Sigma}_1)$. Rewriting the above as second-order
dynamics in $\delta {\bf x}^i$ implies
\begin{eqnarray}
{\bf D}(i) &=& - {\bf Re}^{(+1)} - {\bf Im}^{(+1)} {\bf Re}  {\bf
  Im}^{-1} \nonumber \\
{\bf K}(i) &=& {\bf Im}^{(+1)} {\bf Re} {\bf Im}^{-1} {\bf Re}
+ {\bf Im}^{(+1)} {\bf Im} \nonumber
\end{eqnarray} 

Note that only the change in the time-indices makes this analytic
stability result correct in comparison to a standard LTI approximation
of the LTV system.}{complexLTVdisc}
Let us now consider the placement of real and state-dependent
contraction rates.
\Example{}{Consider the second-order discrete system 
$$
x^{i+2} = f^i(x^{i+1}, x^i, i) + u^i
$$
with position $x^i$ and control input $u^i$. 

Let us now schedule $\sigma_1(z_1^i, i)$ and $\sigma_2(z_2^i,i)$
with $z_1^i=x^i, z_2^i = z_1^{i+1} - \int \sigma_1 dz_1^i$ in the
characteristic equation (\ref{eq:discontrolchar}) of Theorem
\ref{th:discontroller}
\begin{equation}
\left( ^{(+1)} - \int \sigma_2 d \right)
\left( ^{(+1)} - \int \sigma_1 d \right) x^i = 0
\nonumber
\end{equation}
This is equivalent to require the control input $u^i$ in
\begin{eqnarray}
x^{i+2} &=& f^i(x^{i+1}, x^i, i) + u^i \nonumber \\
&=& - \int \sigma_1(z_1^{i+1}, i+1) d z_1^{i+1} + \int
\sigma_2(z_2^i, i) d z_2^i \nonumber
\end{eqnarray}
where the time-varying integration constant can be chosen to achieve
tracking-behaviour of the controller to a desired
trajectory.}{discsecond}
Finally let us consider the placement of complex and state-dependent
contraction rates.
\Example{}{Let us now schedule non-linear complex contraction rates
for a second-order discrete system by requiring the first-order
complex dynamics
\begin{equation}
z^{i+1} = \frac{1}{2} (z^i)^2 + u_d^i(i) \label{eq:zz22}
\end{equation}
with complex contraction rate $\sigma = |z^i|$. In principle any
differentiable complex function can be used here to schedule the
state-dependent complex contraction rates as we want.

The convergence rate of an arbitrary trajectory $z_1$ to another
trajectory $z_2$ is
\begin{equation}
\frac{s^{i+1}}{s^i} = \frac{ \left({\bf z}_1^{i+1} - {\bf z}_2^{i+1}
\right)^{\ast T} \left({\bf z}_1^{i+1} - {\bf z}_2^{i+1} \right) } {
\left({\bf z}_1^i - {\bf z}_2^i \right)^{\ast T} \left({\bf z}_1^i -
  {\bf z}_2^i \right) } = \frac{1}{2} | {\bf z}_1^{i+1} + {\bf
z}_2^{i+1} | \nonumber
\end{equation}
according to (\ref{eq:si1si}) Theorem \ref{th:theoremFdis}. This
region of convergence is naturally larger then the contraction region
$\sigma \le 1$.

The complex dynamics is illustrated in figure \ref{fig:disc}. We can see that $|\sigma|$ increases
from the stable origin. We find excactly two equilibrium points at
$z_1^i= 0$ and $z_2^i = 2$ with constant distance $\frac{s^{i+1}}{s^i}
= 1$. 
\begin{figure}
\begin{center}
\includegraphics[scale=0.4]{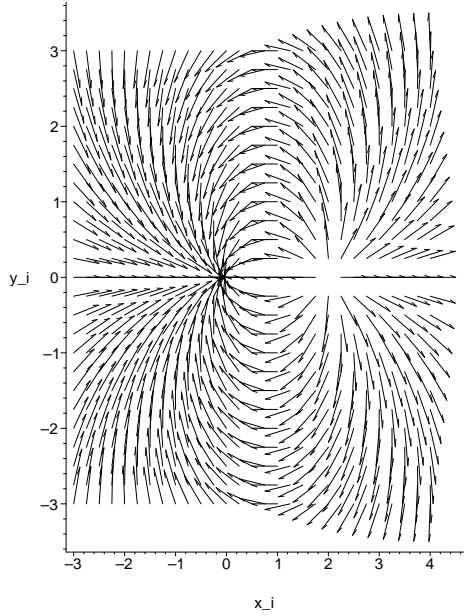}
\end{center}
\caption{Quadratic complex discrete state space dynamics}
\label{fig:disc}
\end{figure}

The complex dynamics is with $x^i = \Re(z^i)$ and $y^i = \Im(z^i)$
equivalent to
\begin{eqnarray}
x^{i+1} &=& \frac{1}{2} (x^i)^2 - \frac{1}{2} (y^i)^2 + u_d^i \nonumber
\\
y^{i+1} &=& x^i y^i \nonumber
\end{eqnarray}
with corresponding Jacobian
$$
\left( \begin{array}{cc} x^i & -y^i \\ y^i & x^i \end{array}
\right) 
$$
that is contracting with $\sigma = \sqrt{ (x^i)^2 + (y^i)^2 }$. 

Hence the corresponding real second-order plant dynamics to
(\ref{eq:zz22}) is
$$
x^{i+2} = \frac{1}{2} (x^{i+1})^2 + x^{i+1} (x^i)^2 - \frac{1}{2}
(x^i)^4 - (x^i)^2 u^i + u^{i+1}
$$
to which the same convergence results apply.
}{complexquadraticdisc}

\section{Discrete-time observers} \label{disobserver}

In this section we consider $\forall i \ge 0$ a smooth $n$-th order
dynamic system in observability form
$$
{\bf x}^{i+n} = {\bf a}_1^{(+n-1)}({\bf x}^i, i) + {\bf
a}_2^{(+n-2)}({\bf x}^i, i) + ... + {\bf a}_n({\bf x}^i, i)
$$
with $M$-dimensional measurement ${\bf y}^i({\bf x}^i, i)$,
$N$-dimensional state ${\bf x}^i$, $N$-dimensional
non-linear plant dynamics ${\bf a}_j({\bf x}^i, i)$ and time index
$i$, which is equivalent to
\begin{equation}
{\bf x}^{i+1}_j = {\bf x}_{j+1}^i + {\bf a}_j({\bf
x}^i, i) \ \mbox{for } j = 1, ..., n \label{eq:displant}
\end{equation}
with ${\bf x}^i_1 = {\bf x}^i$ and $\hat{\bf x}^i_{n+1} = {\bf
0}$.

Let us now introduce the observer 
$$
\hat{\bf x}^{i+1}_j = \hat{\bf x}_{j+1}^i + {\bf a}_j(\hat{\bf
x}^i, i) + {\bf e}_j(\hat{\bf y}^i, i) - {\bf e}_j({\bf y}^i, i)
\ \mbox{for } j = 1, ..., n
$$
with $\hat{\bf x}^i_1 = \hat{\bf x}^i$ and $\hat{\bf x}^i_{n+1} = {\bf
0}$ that allows to extend the plant dynamics ${\bf a}_j$ with a
chosable measurement feedback ${\bf e}_j$ in the equivalent $n$-th
order observer dynamics
\begin{equation}
\hat{\bf x}^{i+n} = \sum_{j=1}^n \left( {\bf a}_j(\hat{\bf x}^i, i) +
{\bf e}_j(\hat{\bf y}^i, i) - {\bf e}_j({\bf y}^i, i) \right)^{(+n-j)}
\label{eq:dishighobs}
\end{equation}

Let us now generalize the well-known LTI eigenvalue-placment in Jordan
form to the placement of the hieracial complex dynamics
\begin{equation}
\hat{\bf z}_j^{i+1} = \int {\bf \Lambda}_j (\hat{\bf z}_j^i, i) d
\hat{\bf z}_j^i + \Re(\hat{\bf z}_{j+1}^i)\ \mbox{for } j = 1, ...,
p-1
\label{eq:dhatzn1}
\end{equation} 
with $\Re(\hat{\bf z}_1^i) = \hat{\bf x}^i, \hat{\bf
z}_{p+1}^i = 0$ and where $p$ is given by $n$ minus the number of complex
contraction rate matrices ${\bf \Sigma}_j$. Taking the variation of
the above implies the time-or state-dependent complex contraction rate
matrices ${\bf \Sigma}_j^i$ in 
\begin{equation}
\delta \hat{\bf z}_j^{i+1} \ = \ {\bf \Sigma}_j^i ( \hat{\bf
z}_j^i, i) \delta \hat{\bf z}_j^i + \delta \hat{\bf z}_{j+1}^i
\ \mbox{for } j = 1, ..., p \nonumber
\end{equation}
According to Theorem \ref{th:theoremFdis} is the stability of this
hierachy given by the largest singular value of ${\bf \Sigma}_j$.

Substituting the $p$ dynamics (\ref{eq:dhatzn1}) recursively in each
other leads with  to
\begin{theorem} 
Given the smooth $n$-th order dynamic system in observability form
\begin{equation}
{\bf x}^{i+n} = {\bf a}_1^{(+n-1)}({\bf x}^i, i) + {\bf
  a}_2^{(+n-2)}({\bf x}^i, i) + ... + {\bf a}_n({\bf x}^i, i)
\label{eq:disobsdynamics}
\end{equation}
with $M$-dimensional measurement ${\bf y}({\bf x}^i, i)$ of the
$N$-dimensional state ${\bf x}^i$, $N$-dimensional
non-linear plant dynamics ${\bf a}_j({\bf x}^i, i)$ and time index
$i$.

An observer 
\begin{equation}
\hat{\bf x}^{i+1}_j = \hat{\bf x}_{j+1}^i + {\bf a}_j(\hat{\bf x}^i,
i) + {\bf e}_j(\hat{\bf y}^i, i) - {\bf e}_j({\bf y}^i, i) \ \mbox{for
} j = 1, ..., n \label{eq:disobserver}
\end{equation}
with $\hat{\bf x}_1^i = \hat{\bf x}^i = \Re(\hat{\bf z}_1)$ and
$\hat{\bf x}^i_{n+1} = {\bf 0}$ allows to place with the measurement
feedback terms ${\bf e}_j$ the time- or state-dependent, integrable,
complex contraction rate matrices ${\bf \Sigma}_j(\hat{\bf z}_j^i, i)$
in the characteristic equation
\begin{equation}
\left( ^{(+1)} - \int \Sigma_p d \right)\Re ...
\left( ^{(+1)} - \int \Sigma_1 d \right) \hat{\bf x}^i = {\bf 0}
\label{eq:disobserverchar}
\end{equation}
with $\Re(\hat{\bf z}_1^i) = \hat{\bf x}^i, \Re(\hat{\bf z}_{j+1}^i) =
\hat{\bf z}_j^{i+1} - \int {\bf \Lambda}_j (\hat{\bf z}_j^i, i) d
\hat{\bf z}_j^i$.

The largest singular value of ${\bf \Sigma}_j^i(\hat{\bf z}_j^i, i)$
implies global contraction behavior of the observer state with ${\bf
\Sigma}_j^i(\hat{\bf z}_j^i, i)$ to the plant state according to
Theorem \ref{th:theoremFdis}.

$p$ is given by $n$ minus the number of complex contraction rate
matrices ${\bf \Sigma}_j^i$ and $\Re$ applies to its left-hand
term. \label{th:disobserver}
\end{theorem}

This theorem generalizes the extended LTV Luenberger observer design
of constant eigenvalues (see e.g. \cite{Lohm1}, \cite{Mracek} or
\cite{Zeitz}) to non-linear or state-dependent contraction rates for
non-linear, time-varying systems.

Note that the global controller in Theorem \ref{th:discontroller} that
uses the state estimates of the global observer in Theorem
\ref{th:disobserver} satisfies a separation principle. Indeed,
subtracting the plant dynamics (\ref{eq:displant}), eventually extended by
a control input ${\bf G}({\bf y}^i, i){\bf u}^i(\hat{\bf x}^i, i)$,
from the observer dynamics (\ref{eq:disobserver}), that is extended by
the same control input ${\bf G}({\bf y}^i, i){\bf u}^i(\hat{\bf x}^i,
i)$, leads with $\tilde{\bf x}^i = \hat{\bf x}^i - {\bf x}^i$ and the
mid-point theorem to
\begin{equation}
\tilde{{\bf x}}_j^{i+1} = \tilde{\bf x}_{j+1}^i - \frac{\partial
\left( {\bf a}_j + {\bf e}_j \right) }{\partial {\bf x}^i} ({\bf \xi},
i) \tilde{\bf x}^i \ \mbox{for } j = 1, ..., n \nonumber
\end{equation}
with $\tilde{\bf x}_{n+1}^i = 0$ and where ${\bf \xi}$ is one point
between ${\bf x}^i$ and $\hat{\bf x}^i$. We can see that the Jacobian
of the error-dynamics of the observer is unchanged. Since
$\frac{\partial {\bf u}^i}{\partial {\bf x}^i}$ in Theorem
\ref{th:contcontroller} is bounded the controller represents a
hierarchical system \cite{Lohm1}. As a result is the convergence rate
of the controller unchanged as well.

Let us now show how a general $N n$ dimensional plant
$$
\underline{\bf x}^{i+1} = {\bf f}(\underline{\bf x}^i, i)
$$
with $N$-dimensional measurement ${\bf y}^i = {\bf x}^i(\underline{\bf
x}^i, i)$ can be transformed to the higher-order observability form
(\ref{eq:disobsdynamics}). A necessary condition is that the mapping
$$
\left( \begin{array}{c} {\bf y}^i(\underline{\bf x}^i, i) \\ \vdots \\
{\bf y}^{i+n-1}(\underline{\bf x}^i, i) \end{array} \right) 
$$
can be inverted to $\underline{\bf x}^i({\bf y}^i, ..., {\bf
  y}^{i+n-1}, i)$ such that we get an explict dynamics
(\ref{eq:disobsdynamics})
$$
{\bf x}^{i+n} = {\bf x}^{i+n}(\underline{\bf x}^i({\bf y}^i, ..., {\bf
  y}^{i+n-1}, i), i)
$$
Hence a necessary (but not sufficientI) observability condition is
that the observability matrix 
$$
{\bf 0} = \left( \begin{array}{c} L^o {\bf c} \\ \vdots \\
L^{n-1} {\bf c} \end{array} \right)  
$$
with the Lie derivatives \cite{Lovelock} $L^o {\bf c} = \frac{\partial
{\bf y}^i}{\partial \underline{\bf x}^i}$ and $L^{j+1} {\bf c} =
\frac{\partial {\bf y}^{i+j+1}}{\partial \underline{\bf x}^i} =L^j
{\bf c}^{(+1)} \frac{\partial {\bf f}}{\partial \underline{\bf
x}}^{(+j)}$ has piece-wisely full rank.  Note that for LTV systems it
is also sufficient.

Let us now consider the observer design of a specific non-linear
problem before we go to the general non-linear case:
\Example{}{Consider the logistic map dynamics 
$$
b^{i+1} = c^i \ b^i \ (1 - b^i) 
$$
with measured state $y^i=b^i$ and unknown constant gain $c^i$. We
can rewrite the above as second-order system
$$
b^{i+2} = \frac{b^{i+1} \ b^{i+1} \ (1 - b^{i+1})}{b^i (1 - b^i)}
$$
Introducing the complex state $x^i = \ln b^i$ we get
$$
x^{i+2} = a_1^{(+1)} + a_2
$$
with $a_1 = \ln ( b^i b^i (1 - b^i) )$ and $a_2 = - \ln ( b^i (1 -
b^i) )$.

We want to build an observer (\ref{eq:disobserver}) with estimate 
$\hat{x}^i = \hat{x}_1^i$
$$
\left( \begin{array}{c} \hat{x}_1^{i+1} \\ \hat{x}_2^{i+1} \end{array}
\right) = \left(
\begin{array}{c} \hat{x}_2^i + a_1(\hat{x}^i) \\ a_2(\hat{x}^i)
\end{array} \right) 
+ 
\left( \begin{array}{c}
e_1(\hat{y}^i) - e_1(y^i) \\
e_2(\hat{y}^i) - e_2(y^i)
\end{array} \right)
$$
with designed constant contraction rates $|\sigma_1|, |\sigma_2| <
1$. Comparing the equivalent second-order dynamics
(\ref{eq:dishighobs}) in $\hat{x}^i$
\begin{equation}
\hat{x}^{i+2} - x^{i+2} = \left( a_1(\hat{x}^i) - a_1(x^i)+
e_1(\hat{x}) - e_1(x^i) \right)^{(+1)} + \left( a_2(\hat{x}^i) -
a_2(x^i)+ e_1(\hat{x}) - e_1(x^i) \right) \nonumber
\end{equation}
with the characteristic equation (\ref{eq:disobserverchar})
\begin{equation}
\left( ^{(+1)} - \int \sigma_2 d \right) \left( ^{(+1)} - \int
\sigma_1 d \right) (\hat{x}^i - x^i) = 0 \nonumber
\end{equation}
leads to the non-linear feedback
\begin{eqnarray}
e_1(\hat{y}^i) + a_1(\hat{y}^i) &=& (\sigma_1 + \sigma_2) \hat{y}^i
\nonumber \\
e_2(\hat{y}^i) + a_2(\hat{y}^i) &=& - \sigma_1 \sigma_2 \hat{y}^i
\nonumber 
\end{eqnarray}

Note that the observer can be transformed back to the real coordinates
$\hat{b}_1^i = e^{\hat{x}_1^i}, \hat{b}_2^i = e^{\hat{x}_2^i}$ as
$$
\left( \begin{array}{c} \hat{b}_1^{i+1} \\ \hat{b}_2^{i+1} \end{array}
\right) = \left(
\begin{array}{c} 
\hat{b}_2^i \ \hat{b}^i_1 \hat{b}^i_1 (1 - \hat{b}^i_1) \
e^{e_1(\hat{y}^i) - e_1(y^i)} \\
\frac{1}{b^i_1 ( 1 - \hat{b}^i_1 )} e^{e_2(\hat{y}^i) - e_2(y^i)}
\end{array} \right) 
$$
such that we can compute $\hat{b}^i = \hat{b}_1^i$ and the estimated
unknown gain as
$$
\hat{c}^i = - \frac{\hat{b}^{i+1}_1}{\hat{b}^i_1 \ (1 - \hat{b}^i_1)}
= \hat{b}_2^i \ \hat{b}^i_1 e^{e_1(\hat{y}^i) - e_1(y^i)} 
$$
}{logmap}
The following example gives an explicit equation for the feedback gains
to achieve time-dependent contraction rates:
\Example{}{Consider the $n$-dimensional non-linear system dynamics
$$
{\bf x}^{i+n} = {\bf a}_1^{(+n-1)}({\bf x}^i, i) + {\bf a}_2^{(+n-2)}
+ ...  + {\bf a}_n({\bf x}^i, i)
$$
with non-linear plant dynamics ${\bf a}_j({\bf x}^i, i)$ and
measurement ${\bf y}^i(x^i,i)$.

Comparing the $n$-th order dynamics (\ref{eq:dishighobs}) of the
observer (\ref{eq:disobserver}) to the characteristic equation
(\ref{eq:disobserverchar}) of real time-varying contraction rates
${\bf \Sigma}_j(i)$ implies the feedback gains
\begin{eqnarray}
{\bf e}_n(\hat{\bf y}^i,i) + \hat{\bf a}_n(\hat{\bf x}^i,i) &=&
\Pi_{j=n}^1 {\bf \Sigma}_j \hat{\bf x}^i \nonumber \\ &\vdots&
\nonumber \\ 
{\bf e}_1(\hat{\bf y}^i,i) + \hat{\bf a}_1(\hat{\bf x}^i,i) &=&
\sum_{j=1}^n {\bf \Sigma}_j^{(-j+1)} \hat{\bf x}^i \nonumber
\end{eqnarray}
}{pseudolineardisobserver}

\section{Discrete higher-order analysis and robustness}
\label{discretehigher}

Consider for $i \ge 0$ the $n$-th dimensional ($n\ge1$) system
$$
{\bf x}^{i+n} = {\bf f}({\bf x}^i, ..., {\bf x}^{i+n-1}, i)
$$
with $N$-dimensional position ${\bf x}^i$. 

In Theorem \ref{th:discontroller} and \ref{th:disobserver} the
characteristic equation of the dynamics above is zero since we use the
observer or controller feedback to precisely achieve the
characteristic equation. For such a given controller or observer an
additional modelling uncertainty ${\bf d}$ may have to be considered
on top to the designed characteristic dynamics. This introduces the
idea of the existence of a perturbation ${\bf d}$ in the
characteristic equation if we analyse a given ODE.

Based on this thought let us approximate this
dynamics with the complex, integrable contraction rates ${\bf
\Sigma}_j^i({\bf z}_j^i, i)$ - that eventually correspond to the designed
contraction rates - in the distorted characteristic equation
\begin{equation}
\left( ^{(+1)} - \int {\bf \Sigma}_p d \right)\Re ...
\left( ^{(+1)} - \int {\bf \Sigma}_1 d \right) {\bf x}^i = {\bf
  d}({\bf z}_p^i, ... {\bf z}_1^i, i) \nonumber
\end{equation}
with $\Re({\bf z}_1) = {\bf x}$ and $\Re({\bf z}_{j+1}^i) = {\bf
z}_j^{i+1} - \int {\bf \Sigma}_j ({\bf z}_j^i, i) d{\bf z}_j^i$. For a
controller or observer of Theorem \ref{th:discontroller} or
\ref{th:disobserver} ${\bf d}$ may represent the modelling
uncertainities of the system.

Taking the variation of the above we get
\begin{equation}
\delta {\bf z}_p^{i+1} - {\bf \Sigma}_p \delta {\bf z}_p^i =
\frac{\partial {\bf d}}{\partial {\bf z}_p^i} \delta {\bf z}_p^i + \
... \ + \frac{\partial {\bf d}}{\partial {\bf z}_1^i} \delta {\bf
z}_1^i \nonumber
\end{equation}
The main idea is to construct an exponential bound on the virtual
displacement $\delta {\bf x}^i$ over $p$ time-steps, rather than over
a single time-step as in \cite{Lohm1}.

Let us first bound the higher-order term by taking the norm of the
above
\begin{equation}
|\delta {\bf z}_p^{i+1} - {\bf \Sigma}_p \delta {\bf x}_p^i| \le
|\frac{\partial {\bf d}}{\partial {\bf z}_p^i}| |\delta {\bf z}_p^i| +
\ldots + |\frac{\partial {\bf d}}{\partial {\bf z}_1^i}| |\delta {\bf
z}_1^i| \label{eq:disnormdyn}
\end{equation}
where now and in the following the norm of a matrix is the largest
singular value of that matrix and the norm of a vector is the
root of the vector multiplied with its conjungate vector.

Let us now select a real $\eta^i({\bf z}_1^i, ..., {\bf z}_p^i, i) \ge
0$ that fulfils 
\begin{equation} 
\eta^{i+p-1}...\eta^i \ge |\frac{\partial {\bf d}}{\partial {\bf
z}_p^i}| \eta^{i+p-2}...\eta^i + \ \ldots \ + |\frac{\partial {\bf
d}}{\partial {\bf z}_1^i}| \label{eq:characteristicdis}
\end{equation}
$\forall i \ge 0$. Let us bound the initial conditions at $i=0$ with
real and constant $K \ge 0$ as
\begin{equation} 
|\delta {\bf z}_j^i| \le K \ \eta^{i+j-2}...\eta^i \ \Pi_{k=0}^{i-1}
(\eta^k + \sigma_{max}^k), \ 1 \le j \le p \label{eq:dxndis}
\end{equation}
where $\sigma_{max}^i({\bf z}_1^i, ..., {\bf z}_p^i, i)$ is the
largest singular value of all ${\bf \Sigma}_j({\bf z}_j^i, i)$. Hence
with (\ref{eq:characteristicdis}) and (\ref{eq:dxndis}) we can bound
(\ref{eq:disnormdyn}) at $i=0$ as
\begin{eqnarray}
|\delta {\bf z}_p^{i+1} - {\bf \Sigma}_p \delta {\bf z}_{n-1}^i|
&\le& K \ \Pi_{k=0}^{i-1} (\eta^k + \sigma_{max}^k) \left(
|\frac{\partial {\bf d}}{\partial {\bf z}_p^i}| \eta^{i+p-2}...\eta^i
+ \ \ldots \ + |\frac{\partial {\bf d}}{\partial {\bf z}_1^i}| \right)
\nonumber \\ &\le& K \ \eta^{i+p-1}...\eta^i \ \Pi_{k=0}^{i-1} (\eta^k
+ \sigma_{max}^k) \label{eq:dxn1dis}
\end{eqnarray}
Theorem \ref{th:theoremFdis} on $\delta {\bf z}_j^{i+1} = \Lambda_j +
\Re({\bf z}_{j+1})$ with the bounded distortion (\ref{eq:dxndis}) and
(\ref{eq:dxn1dis}) implies at $i=1$
\begin{eqnarray}
|\delta {\bf z}^{i+1}_j| &\le& \sigma_{max}^i |\delta {\bf z}^i_j| +
K \ \eta^{i+j-1}...\eta^i \ \Pi_{k=0}^{i-1} (\eta^k + \sigma_{max}^k)
\nonumber \\
&\le& K \ \eta^{i+j-1}...\eta^{i+1} \ \Pi_{k=0}^i (\eta^k +
\sigma_{max}^k), 1 \le j \le p \nonumber
\end{eqnarray}
which implies with complete induction that (\ref{eq:dxndis}) and
(\ref{eq:dxn1dis}) hold $\forall i \ge 0$.

Using the above this allows to conclude:
\begin{theorem}
Consider for $i \ge 0$ the $n$-dimensional ($n \ge 1$) system
$$
{\bf x}^{i+n} = {\bf f}^i({\bf x}^i, ..., {\bf x}^{i+n-1}, i)
$$
with $N$-dimensional position ${\bf x}^i$ at time $i$.

Let us approximate the above dynamics with the integrable, complex
contraction rates ${\bf \Sigma}_j({\bf z}_j^i, i)$ in the distorted
characteristic equation
\begin{equation}
\left( ^{(+1)} - \int {\bf \Sigma}_p d \right)\Re ...
\left( ^{(+1)} - \int {\bf \Sigma}_1 d \right) {\bf x}^i = {\bf
  d}({\bf z}_p^i, ... {\bf z}_1^i, i) \label{eq:dischardist}
\end{equation}
with $\Re({\bf z}_1^i) = {\bf x}^i$ and $\Re({\bf z}_{j+1}^i) = {\bf
z}_j^{i+1} - \int {\bf \Sigma}_j ({\bf z}_j^i, i) d{\bf z}_j^i$.

Bounding the effect of the distortion ${\bf d}$ with a real
$\eta^i({\bf z}_1^i,...,{\bf z}_p^i, i) \ge 0$ that fulfils $\forall
i \ge 0$
\begin{equation} 
\eta^{i+n-1}...\eta^i \ge |\frac{\partial {\bf d}}{\partial {\bf z}_p}|
\eta^{i+n-2}...\eta^i + \ \ldots \ + |\frac{\partial {\bf d}}{\partial
  {\bf z}_1}| \label{eq:discharacteristic}
\end{equation}
and the largest singular value $\sigma_{max}^i({\bf z}_1^i,...,{\bf
  z}_p^i, i)$ of all ${\bf \Sigma}_j({\bf z}_j^i, i)$ leads to global
contraction behaviour with
\begin{equation}
{\bf \Sigma}_j({\bf z}_j^i, i) + \eta^i \label{eq:higherdisrate}
\end{equation}
according to Theorem \ref{th:theoremFdis}  where initial overshoots are
bounded $\forall i \ge 0$ with constant $K > 0$ by
\begin{equation}  
|\delta {\bf z}_j^i | \le K \ \eta^{j-1}...\eta^o(i=0) \ \ (\eta^{i-1}
+ \sigma_{\max j}^{i-1}) ... (\eta^o + \sigma_{\max j}^0), \ 1 \le j
\le p \label{eq:disdefinitionofK} 
\end{equation}

$p$ is given by $n$ minus the number of complex contraction rate
matrices ${\bf \Sigma}_j$ and $\Re$ applies to its left-hand term.
\label{th:higherorderdis}
\end{theorem}

\begin{figure}
\begin{picture}(200,200)(0,-100)
\put(-2,0){\vector(1,0){210}} \put(3,-100){\vector(0,1){200}}
\put(215,0){$i$} \put(0,105){$|\delta z_j^i|$} \put(-10,0){$0$}
\put(25,-2){$:$} \put(50,-2){$:$} \put(75,-2){$:$} \put(100,-2){$:$}
\put(125,-2){$:$} \put(150,-2){$:$} \put(175,-2){$:$}
\put(0,30){$\bullet$} \put(25,-30){$\bullet$} \put(50,40){$\bullet$}
\put(75,-35){$\bullet$} \put(100,-15){$\bullet$}
\put(125,-28){$\bullet$} \put(150,17){$\bullet$}
\put(175,-15){$\bullet$} \qbezier(0,90)(100,27)(200,13)
\qbezier(0,-90)(100,-27)(200,-13)
\end{picture}
\caption{Bound on $\delta z_j^i$ over $i$}
\label{fig:discretecontraction}
\end{figure}
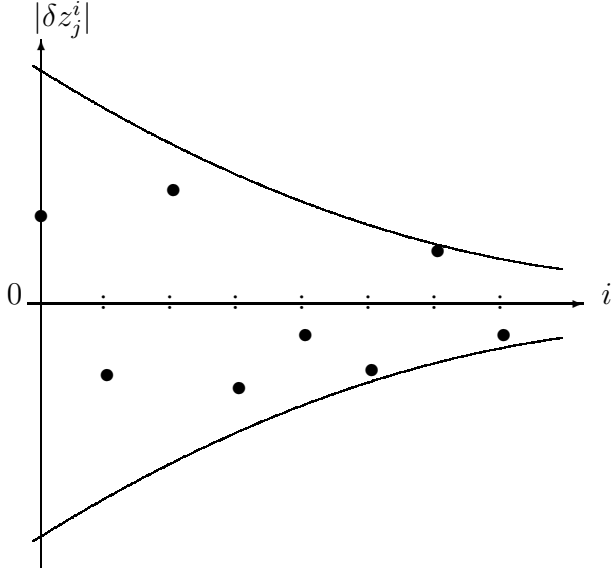

Figure~\ref{fig:discretecontraction} illustrates the exponential bound
(\ref{eq:dxndis}) which allows short-term over-shoots but implies
exponential convergence on the long-term.

Interpreting ${\bf \Sigma}_j$ as the desired contraction rates in
Theorem \ref{th:discontroller} or \ref{th:disobserver} allows to bound
the potential instabilities which come from modelling uncertainties of
the plant.  I.e. it allows to prove robustness for modelling
uncertainties with a bounded de-stabilizing divergence rate $\eta^i$.
Note that additional time-varying errors in the control input do not
affect the contraction behaviour, but the desired trajectory ${\bf
x}_d^i(i)$.

If we cannot design ${\bf \Sigma}_j({\bf z}_j^i, i)$ per feedback then
we have to approximate ${\bf \Sigma}_j$ to minimize the distortion
${\bf d}$, e.g. by transforming the higher-order system in its reduced
form (i.e. a form in which ${\bf f}$ is independent of ${\bf
x}^{i+n-1}$). This is illustrated in the following examples:
\Example{}{In economics, consider the price dynamics
\begin{eqnarray}
{\bf n}^{i+1} &=& {\bf f}({\bf p}^i, i) \nonumber \\
{\bf p}^{i+1} &=& {\bf g}({\bf n}^i, i) \nonumber
\end{eqnarray}
with ${\bf n}^i$ the number of sold products at time $i$ and
corresponding price ${\bf p}^i$. 

The first line above defines the customer demand as a reaction to a
given price. The second line defines the price, given by the
production cost under competition, as a reaction to the number of sold
items.  The dynamics above corresponds to the second-order economic
growth cycle dynamics
$$
{\bf n}^{i+2} = {\bf f} \left( {\bf g} ({\bf n}^i, i) \right)
$$

Contraction behaviour of this economic behaviour with contraction rate
$\eta^i$ can then be concluded with equation
(\ref{eq:discharacteristic}) in Theorem \ref{th:higherorderdis} for
${\bf \Sigma}_1 = {\bf \Sigma}_2 = {\bf 0}$
\begin{equation}
\eta^{i+1} \eta^i \ge \left| \frac{\partial {\bf f}}{\partial {\bf
p}^i} \frac{\partial {\bf g}}{\partial {\bf n}^i} \right|
\label{eq:game}
\end{equation}
That means we get stable (contraction) behaviour if the product of
customer demand sensitivity to price and production cost sensitivity
to number of sold items has singular values less than 1. We get
unstable (diverging) behaviour for the opposite case. 

Note that this result even holds when no precise model of the
sensitivity is known, which is usually the case in economic or game
situations. Whereas the above is well known for LTI economic models we
can see that the economic behaviour is unchanged for a non-linear,
time-varying economic environment.

The above also corresponds to a game situation (see e.g. \cite{Shamma}
or \cite{Bryson}) between two players with strategic action ${\bf
p}^i$ and ${\bf n}^i$. Both players optimize their reaction ${\bf g}$
and ${\bf f}$ with respect to the opponent's action. We can then again
conclude for (\ref{eq:game}) to global contraction behaviour to a
unique time-dependent trajectory (in the autonomous case, the Nash
equilibrium).}{economics}

\Example{}{Consider the general dynamics
$$
{\bf x}^{i+2} + \frac{\partial U}{\partial {\bf x}^i}^{(+1)} +
\frac{\partial^2 U}{\partial ({\bf x}^i)^2}_{\min} {\bf x}^{i+1} +
\frac{\partial V}{\partial {\bf x}^i} = {\bf 0}
$$
with potentials $U({\bf x}^i, i), V({\bf x}^i, i)$,
$N$-dimensional position ${\bf x}^i$ and where we assume without loss
of generality that the singular values of $\frac{\partial^2
U}{\partial ({\bf x}^i)^2}_{\min}$ correspond $\forall {\bf x}^i$
at a given $i$ to the minimal singular values of $\frac{\partial^2
U}{\partial ({\bf x}^i)^2}$. The virtual dynamics is
$$
\delta {\bf x}^{i+2} + \left( \frac{\partial^2 U}{\partial ({\bf
x}^i)^2}^{(+1)} + \frac{\partial^2 U}{\partial ({\bf x}^i)^2}_{\min}
\right) \delta {\bf x}^{i+1} + \frac{\partial^2 V}{\partial ({\bf
x}^i)^2} \delta {\bf x}^i = {\bf 0}
$$
The distorted characteristic equation (\ref{eq:dischardist}) is
with ${\bf \Sigma}_1 = - \frac{\partial^2 U}{\partial ({\bf
x}^i)^2}$, ${\bf \Sigma}_2 = -\frac{\partial^2 U}{\partial
({\bf x}^i)^2}_{\min}$
\begin{equation}
\left( ^{(+1)} - \Sigma_2 \right) ...  \left( ^{(+1)} - \Sigma_1
\right) \delta {\bf x}^i = \left( \frac{\partial^2 U}{\partial ({\bf
x}^i)^2}_{\min} \frac{\partial^2 U}{\partial ({\bf x}^i)^2}^{(+1)} -
\frac{\partial^2 V}{\partial ({\bf x}^i)^2} \right) \delta {\bf x}^i
\nonumber
\end{equation}
The remaining instability $\eta^i$ in (\ref{eq:discharacteristic}) is
then given by
$$
\eta^{i+1} \eta^i \ge \left| \frac{\partial^2 U}{\partial ({\bf
x}^i)^2}_{\min} \frac{\partial^2 U}{\partial ({\bf x}^i)^2}^{(+1)} -
\frac{\partial^2 V}{\partial ({\bf x}^i)^2} \right|
$$ 
We can hence bound with Theorem \ref{th:higherorderdis} the
contraction rates with 
$$
\frac{\partial^2 U}{\partial (x^{i})^2}_{\min} + \eta^i \le
\frac{\partial^2 U}{\partial (x^{i})^2}_{\min} + \sqrt{
\left| \frac{\partial^2 U}{\partial ({\bf
x}^i)^2}_{\min} \frac{\partial^2 U}{\partial ({\bf x}^i)^2}^{(+1)} -
\frac{\partial^2 V}{\partial ({\bf x}^i)^2} \right| }_{max}
$$
For the scalar case with constant damping this condition is equivalent
to require that the complex poles $\frac{\partial^2 U}{\partial
(x^i)^2} \pm \sqrt{ \frac{\partial^2 U}{\partial (x^i)^2}^2 -
\frac{\partial^2 V}{\partial (x^i)^2}(x^i,i) }$ lie $\forall x^i, i$
within the green square of the complex plane in figure
\ref{fig:discretecirecle}.
\begin{figure}[h]
\begin{center}
\epsfig{figure=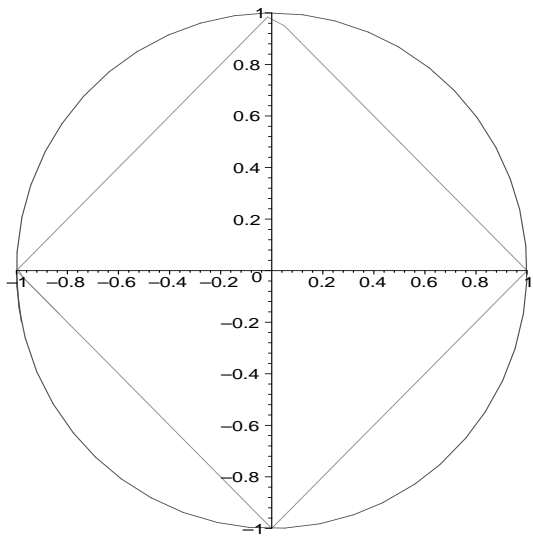,height=70mm,width=70mm}
\end{center}
\caption{LTI stability circle and non-linear contraction square in
complex plane} \label{fig:discretecirecle}
\end{figure}}{discnonlinearsecondorder}

\Example{}{Consider the 2D lighthouse problem in figure \ref{fig:sin}
of navigating a vehicle using only azimuth measurements $y^i$ to a
fixed point in space. The dynamic equations of the vehicle's motion
are
\begin{equation}
{\bf x}^{i+1} = {\bf x}^i + {\bf u}^i \nonumber
\end{equation}
with 2D position ${\bf x}^i = (x_1^i, x_2^i)^T$ and control input
${\bf u}^i = (u_1^i, u_2^i)^T$.  The vehicle measures only the azimuth
to the lighthouse, $\ y^i \ = \ \tan \psi^i \ = \ \frac{x_1^i}{x_2^i}\
$.
\begin{figure}
\begin{center}
\includegraphics[scale=0.4]{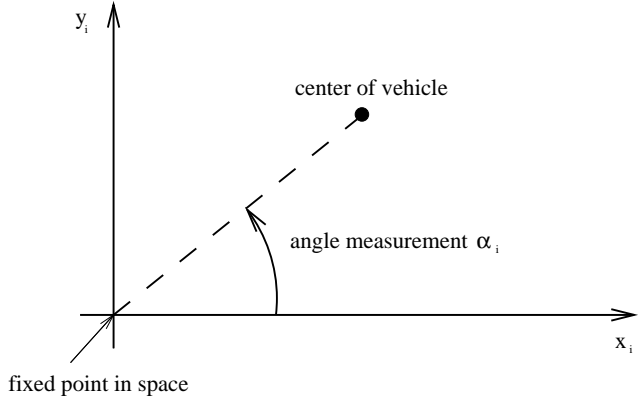}
\end{center}
\caption{Lighthouse navigation}
\label{fig:lighthouse}
\end{figure}

Consider now the observer
\begin{equation}
\hat{\bf x}^{i+1} = \hat{\bf x}^i + {\bf u}^i + (\gamma^i -1)\left(
\begin{array}{c} \cos \psi^i \\ -\sin \psi^i \end{array} \right)
\left(
\begin{array}{cc} \cos \psi^i & -\sin \psi^i \end{array} \right)
\hat{\bf x}^i \label{eq:lighthouseobs}
\end{equation}
From Theorem \ref{th:theoremFdis}, this observer is semi-contracting
for $-1 \le \gamma^i \le 1$. Since the true dynamics is a particular
solution of the observer dynamics we can then conclude on global
convergence of $\hat{\bf x}^i$ to ${\bf x}$.

In the case of no model or measurement uncertainty, the optimal choice
of $\gamma^i$ is 0. Otherwise, the choice of $\gamma^i$ should
trade-off the effect of these uncertainties, as e.g. in the
contraction-based strap-down observer of~\cite{inertial}.

Let us now compute the Jacobian of (\ref{eq:lighthouseobs})
$$
\frac{\partial {\bf f}^i}{\partial {\bf x}^i} = \left(
\begin{array}{cc} \cos \psi^i & \sin
\psi^i \\ -\sin \psi^i & \cos \psi^i \end{array} \right) \left(
\begin{array}{cc} \gamma^i & 0 \\ 0 & 1 \end{array} \right) \left(
\begin{array}{cc} \cos \psi^i & -\sin \psi^i \\ \sin \psi^i & \cos
\psi^i \\ \end{array} \right)
$$
We can compute for constant $\gamma^i = \gamma$ e.g. with MAPLE the
square of the largest singular value of $\frac{\partial {\bf
f}^{i+1}}{\partial {\bf x}^{i+1}} \frac{\partial {\bf f}^i}{\partial
{\bf x}^i}$ as
\begin{eqnarray}
&& \frac{1}{2} (\gamma^4 + 1) \cos(\psi^{i+1} - \psi^{i})^2 +
\gamma^2 \sin(\psi^{i+1} - \psi^{i})^2 \nonumber \\
&&+ \frac{1}{2} |\cos(\psi^{i+1} -
\psi^{i}) (\gamma^2 - 1)| | \cos(\psi^{i+1} - \psi^{i})^2
(\gamma^2-1)^2 + 4 \gamma^2) |^{-\frac{1}{2}} \label{eq:singvals}
\end{eqnarray}  
which simplifies for $\gamma = 0$ to $| \cos(\psi^{i+1} - \psi^{i})
|^2$. Using equation (\ref{eq:discharacteristic}) in Theorem
\ref{th:higherorderdis} for $\sigma_1^i = \sigma_2^i = 0$ the
exponential contraction rate is for $\gamma = 0$
\begin{equation}
\eta^{i+1} \eta^i \ge | \cos(\psi^{i+1} - \psi^{i}) | \label{eq:rate}
\end{equation}  
Thus, we can conclude on contraction behaviour over several measurement
updates if $\psi^i$ changes over different $i$.

Let us now illustrate the above results with simple simulations in the
2D case with position $x_1^i$ and $x_2^i$ over the time index $i$ with
measurement $y^i$.

Figure \ref{fig:tangent} shows the motion of a vehicle with constant
velocity vector, which is initially tangential to the lighthouse. Due
to the tangential motion leads the observer (\ref{eq:lighthouseobs})
with $\gamma = 0$ to global exponential convergence to the real
trajectory with convergence rate (\ref{eq:rate}).
\begin{figure}
\begin{center}
\includegraphics[scale=0.4]{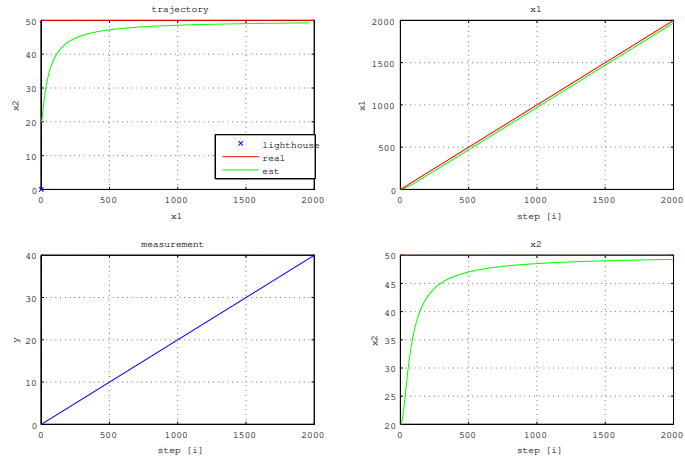}
\end{center}
\caption{Tangential movement with respect to lighthouse}
\label{fig:tangent}
\end{figure}

Figure \ref{fig:radial} shows a vehicle with constant velocity vector
radial to the lighthouse. The observer (\ref{eq:lighthouseobs}) with
$\gamma = 1$ achieves global semi-contraction behaviour, i.e. the
tangential error disappears, whereas the non-observable radial error
remains.
\begin{figure}
\begin{center}
\includegraphics[scale=0.4]{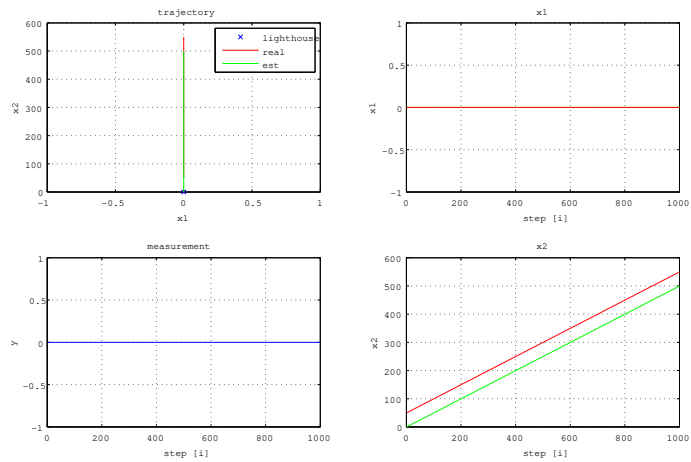}
\end{center}
\caption{Radial movement with respect to lighthouse}
\label{fig:radial}
\end{figure}

Consider now the 3D lighthouse problem of navigating a vehicle
using only azimuth $\psi^i$ and elevation measurements $\theta^i$ to a
fixed point in space. The position dynamics of the vehicle's motion is
\begin{eqnarray}
{\bf x}^{i+1} &=& {\bf x}^i + {\bf u}^i \nonumber
\end{eqnarray}
with 3D position ${\bf x}^i = (x_1^i, x_2^i, x_3^i )^T$ and control
input ${\bf u}^i = (u_1^i, u_2^i, u_3^i)^T$. The vehicle measures only
azimuth $\ y_1^i\ = \ \tan \psi^i \ = \ \frac{x_1^i}{x_2^i}\ $ and
elevation $\ y_2^i\ = \ \tan \theta^i = \frac{x_3^i}{\sqrt{x_1^i
x_1^i + x_2^i x_2^i}}\ $ to the lighthouse. 

The measurement equations can be rewritten in a LTV
form in ${\bf x}^i$ as
\begin{eqnarray}
x_2^i \tan \psi^i - x_1^i &=& 0 \nonumber \\
x_2^i \tan \xi^i - x_3^i &=& 0 \nonumber \\
x_1^i \tan \zeta^i - x_3^i &=& 0 \nonumber
\end{eqnarray}
with $\tan \xi^i = \frac{\tan \theta^i}{\cos \psi^i}$ and $\tan
\zeta^i = \frac{\tan \psi^i \tan \theta^i}{\cos \psi^i}$.  

Consider now the observer 
\begin{eqnarray}
\hat{\bf x}^{i+1} = \hat{\bf x}^i + {\bf u}^i &+& (a^i(i)-1) \left(
\begin{array}{c} \cos \psi^i \\ -\sin \psi^i \\ 0 \end{array} \right)
\left( \begin{array}{ccc} \cos \psi^i & -sin \psi^i & 0 \end{array}
\right) \hat{\bf x}^i \nonumber \\
&+& (b^i(i)-1)
\left( \begin{array}{c} 0 \\ -\sin \xi^i \\ \cos \xi^i \end{array}
\right) \left( \begin{array}{ccc} 0 & -\sin \xi^i & \cos \xi^i
\end{array} \right) \hat{\bf x}^i \nonumber \\
&+& (c^i(i)-1)
\left( \begin{array}{c} -\sin \zeta^i \\ 0 \\ \cos \zeta^i \end{array}
\right) \left( \begin{array}{ccc} -\sin \zeta^i & 0 & \cos \zeta^i
\end{array} \right) \hat{\bf x}^i \nonumber
\end{eqnarray}

This dynamics is a superposition of the 2D-lighthouse problem. Hence
we can conclude horizontally with (\ref{eq:singvals}) or
(\ref{eq:rate}) on exponential convergence over several measurement
updates if $\psi^i$ changes over different $i$. The
vertical exponential convergence rate is then given by the minimum of
$|b^i|$ or $|c^i|$.

Since the true dynamics is a particular solution of the observer
dynamics we can then conclude on global exponential convergence of
$\hat{\bf x}^i$ to ${\bf x}$.}{lighthouseexample}

\section{Concluding Remarks} \label{Conclusion}

This paper derives, for non-linear time-varying systems in
controllability form
\begin{eqnarray}
{\bf x}^{(n)} &=& {\bf f}({\bf x}, ... {\bf x}^{(n-1)}, t) \nonumber \\
{\bf x}^{i+n} &=& {\bf f}({\bf x}^i, ... {\bf x}^{i+n-1}, i) \nonumber
\end{eqnarray}
simple controller designs in Theorem \ref{th:contcontroller} and
\ref{th:discontroller} to achieve specified exponential {\it state-and
time-dependent} convergence rates.  The approach can also be regarded
as a general gain-scheduling technique with global exponential
stability guarantees. The resulting design is illustrated for real and
complex time- and state-dependent contraction rates, inclusive the
inverted pendulum.

A dual observer design technique is also derived for non-linear
time-varying systems in observability form
\begin{eqnarray}
{\bf x}^{(n)} &=& {\bf a}_1^{(n-1)}({\bf x}, t) + {\bf
a}_2^{(n-2)}({\bf x}, t) + ...  + {\bf a}_n({\bf x}, t) \nonumber \\
{\bf x}^{i+n} &=& {\bf a}_1^{(+n-1)}({\bf x}^i, i) + {\bf
a}_2^{(+n-2)}({\bf x}^i, i) + ... + {\bf a}_n({\bf x}^i, i) \nonumber 
\end{eqnarray}
, where so far straightforward observer techniques were not known. The
resulting observer design is illustrated for non-linear chemical
plants, the Van-der-Pol oscillator, the discrete logarithmic map series
prediction and lighthouse navigation problem.

These results allow one to shape state- and time-dependent global
exponential convergence rates ${\bf \Lambda}_j({\bf z}_j, t)$ and
${\bf \Sigma}_j({\bf z}_j, t)$ ideally suited to the non-linear or
time-varying system with the generalized characteristic equation
\begin{eqnarray}
\left(\frac{d}{dt} - \int {\bf \Lambda}_p d \right)\Re ...
\left(\frac{d}{dt} - \int {\bf \Lambda}_1 d \right) {\bf x} &=& {\bf
  0} \nonumber \\ 
\left( ^{(+1)} - \int {\bf \Sigma}_p d \right)\Re ...  \left( ^{(+1)}
- \int {\bf \Sigma}_1 d \right) {\bf x}^i &=& {\bf 0} \nonumber
\end{eqnarray}
with $\Re({\bf z}_1) = {\bf x}, \Re({\bf z}_{j+1}) = \dot{\bf z}_j -
\int {\bf \Lambda}_j ({\bf z}_j, t) d{\bf z}_j$ ( $\Re({\bf z}_1^i)
= {\bf x}^i, \Re({\bf z}_{j+1}^i) = {\bf z}_j^{i+1} - \int {\bf
\Sigma}_j ({\bf z}_j^i, i) d{\bf z}_j^i$ ).

Analytic exponential robustness bounds on general non-linear,
time-varying distortions ${\bf d}$ on the right-hand side of the
characteristic equation are given with $\eta$ in Theorem
\ref{th:higherordercontinuous} or \ref{th:higherorderdis}. Both
theorems can also be used to derive analytic state- and time-dependent
approximated convergence rates for given general non-linear,
time-varying higher-order systems.

Note that the general technique of this paper matches the eigenvalue
analysis for LTI systems. For non-LTI systems additional time
derivatives for continuous systems and index changes for discrete
systems of the contraction rates have to be considered. Only with
these changes exponential convergence guarantees with the time- and
state-dependent contraction rates are given.

\end{document}

%% file: eps.tex
\newcommand{\epsfig}[3]{
  \begin{figure}[htb] 
     \begin{center}
       \begin{minipage}[t]{#2}
         \epsfxsize=#2
         \leavevmode
         \epsfbox{#1}
       \end{minipage}
     \end{center}
     \caption{#3}
     \label{f#1}
   \end{figure}
}
